\documentstyle[epsfig]{mn}

\newif\ifAMStwofonts

\ifoldfss
  \ifCUPmtlplainloaded \else
    \NewTextAlphabet{textbfit} {cmbxti10} {}
    \NewTextAlphabet{textbfss} {cmssbx10} {}
    \NewMathAlphabet{mathbfit} {cmbxti10} {} 
    \NewMathAlphabet{mathbfss} {cmssbx10} {} 
  \fi
  \ifAMStwofonts
    \ifCUPmtlplainloaded \else
      \NewSymbolFont{upmath} {eurm10}
      \NewSymbolFont{AMSa} {msam10} 
      \NewMathSymbol{\upi}     {0}{upmath}{19}
      \NewMathSymbol{\umu}     {0}{upmath}{16}
      \NewMathSymbol{\upartial}{0}{upmath}{40}
      \NewMathSymbol{\leqslant}{3}{AMSa}{36}
      \NewMathSymbol{\geqslant}{3}{AMSa}{3E}

      \let\leq=\leqslant \let\le=\leqslant
      \let\geq=\geqslant 
    \fi
  \fi
\fi 

\ifnfssone
  \newmathalphabet{\mathit}
  \addtoversion{normal}{\mathit}{cmr}{m}{it}
  \addtoversion{bold}{\mathit}{cmr}{bx}{it}
  \newmathalphabet{\mathbfit} 
  \addtoversion{normal}{\mathbfit}{cmr}{bx}{it}
  \addtoversion{bold}{\mathbfit}{cmr}{bx}{it}
  \newmathalphabet{\mathbfss} 
  \addtoversion{normal}{\mathbfss}{cmss}{bx}{n}
  \addtoversion{bold}{\mathbfss}{cmss}{bx}{n}
  \ifAMStwofonts
    \ifCUPmtlplainloaded \else
      \UseAMStwoboldmath
      \makeatletter
      \new@mathgroup\upmath@group
      \define@mathgroup\mv@normal\upmath@group{eur}{m}{n}
      \define@mathgroup\mv@bold\upmath@group{eur}{b}{n}
      \edef\UPM{\hexnumber\upmath@group}
      \new@mathgroup\amsa@group
      \define@mathgroup\mv@normal\amsa@group{msa}{m}{n}
      \define@mathgroup\mv@bold\amsa@group{msa}{m}{n}
      \edef\AMSa{\hexnumber\amsa@group}
      \makeatother
      \mathchardef\upi="0\UPM19
      \mathchardef\umu="0\UPM16
      \mathchardef\upartial="0\UPM40
      \mathchardef\leqslant="3\AMSa36
      \mathchardef\geqslant="3\AMSa3E

      \let\leq=\leqslant \let\le=\leqslant
      \let\geq=\geqslant 
    \fi
  \fi
\fi 
\newcommand{\lya}{Lyman $\alpha$~}
\ifnfsstwo
  \DeclareMathAlphabet{\mathbfit}{OT1}{cmr}{bx}{it}
  \SetMathAlphabet\mathbfit{bold}{OT1}{cmr}{bx}{it}
  \DeclareMathAlphabet{\mathbfss}{OT1}{cmss}{bx}{n}
  \SetMathAlphabet\mathbfss{bold}{OT1}{cmss}{bx}{n}
  \ifAMStwofonts
    \ifCUPmtlplainloaded \else
      \DeclareSymbolFont{UPM}{U}{eur}{m}{n}
      \SetSymbolFont{UPM}{bold}{U}{eur}{b}{n}
      \DeclareSymbolFont{AMSa}{U}{msa}{m}{n}
      \DeclareMathSymbol{\upi}{0}{UPM}{"19}
      \DeclareMathSymbol{\umu}{0}{UPM}{"16}
      \DeclareMathSymbol{\upartial}{0}{UPM}{"40}
      \DeclareMathSymbol{\leqslant}{3}{AMSa}{"36}
      \DeclareMathSymbol{\geqslant}{3}{AMSa}{"3E}

      \let\leq=\leqslant \let\le=\leqslant
      \let\geq=\geqslant 
    \fi
  \fi
\fi 

\ifCUPmtlplainloaded \else
  \ifAMStwofonts \else 
    \def\upi{\pi}
    \def\umu{\mu}
    \def\upartial{\partial}
  \fi
\fi

\title{The Lyman $\alpha$ forest flux probability distribution at ${\boldmath z>3}$\thanks{Based on observations collected at the  European  Southern  Observatory  Very  Large  Telescope,  Cerro Paranal,  Chile -- Programs 077.A-0166(A), 075.A-0464(A), 073.B-0787(A), 166.A-0106(A), 65.O-0296(A).}}
  \author[F. Calura et al.]  {F. Calura$^{1,2,3}$\thanks{E-mail:
      fcalura@oabo.inaf.it}, E. Tescari$^{4,5}$, V. D'Odorico$^{2}$,
    M. Viel$^{2,6}$, S. Cristiani$^{2,6}$, T.-S. Kim$^{7,8}$,
    \newauthor J. S. Bolton$^{9}$\\ (1) INAF, Osservatorio Astronomico
    di Bologna, via Ranzani 1, 40127 Bologna, Italy\\ (2) INAF,
    Osservatorio Astronomico di Trieste, Via G.B. Tiepolo 11, 34131
    Trieste, Italy\\ (3) Jeremiah Horrocks Institute for Astrophysics
    and Supercomputing, University of Central Lancashire, Preston PR1
    2HE, UK \\ (4) Service d'Astrophysique, CEA-Saclay, Orme des
    Merisiers, 91191 Gif-sur-Yvette, France\\ (5) Dipartimento di
    Fisica - Sezione di Astronomia, Universit\`a di Trieste, Via
    G. B. Tiepolo 11, 34131 Trieste, Italy\\ (6) INFN/National
    Institute for Nuclear Physics, Via Valerio 2, I-34127 Trieste,
    Italy\\ (7) Astrophysikalisches Institut Potsdam, An der
    Sternwarte 16, D-14482 Potsdam, Germany\\ (8) Department of
    Astronomy, University of Wisconsin-Madison, 475 N. Charter St.,
    Madison, WI53706, USA \\ (9) School of Physics, University of
    Melbourne, Parkville, VIC 3010, Australia }
	
\pagerange{\pageref{firstpage}--\pageref{lastpage}}
\pubyear{---}

\begin{document}

\maketitle

\label{firstpage}
\begin{abstract}
We present a measurement of the \lya flux probability distribution
function (PDF) obtained from a set of eight high resolution quasar
spectra with emission redshifts at $3.3 \leq z \leq 3.8$.  We
carefully study the effect of metal absorption lines on the shape of
the PDF.  Metals have a larger impact on the PDF measurements at lower
redshift, where there are relatively fewer Lyman $\alpha$ absorption lines.  
This
may be explained by 
an increase in the number of metal lines which are
blended with Lyman $\alpha$ absorption lines toward higher redshift,
but may also be due to the presence of fewer metals in the
intergalactic medium at earlier times.  We also provide a new
measurement of the redshift evolution of the effective optical depth,
$\tau_{\rm eff}$, at $2.8 \leq z\leq 3.6$, and find no evidence for a
deviation from a power law evolution in the $\log(\tau_{\rm
  eff})-\log(1+z)$ plane.  The flux PDF measurements are furthermore
of interest for studies of the thermal state of the intergalactic
medium (IGM) at $z\simeq 3$.  By comparing the PDF to state-of-the-art
cosmological hydrodynamical simulations, we place constraints on the
temperature of the IGM and compare our results with previous
measurements of the PDF at lower redshift.  At redshift $z=3$, our new
PDF measurements are consistent with an isothermal temperature-density
relation, $T=T_{0}\Delta^{\gamma-1}$, with a temperature at the mean
density of $T_{0} = 19250 \pm 4800$ K and a slope
$\gamma=0.90\pm0.21$(1$\sigma$ uncertainties).  In comparison, joint
constraints with existing lower redshift PDF measurements at $z<3$
favour an inverted temperature-density relation with $T_0=17900 \pm
3500$ K and $\gamma=0.70\pm0.12$, in broad agreement with
  previous analyses.

\end{abstract} 


\section{Introduction} 
The Lyman $\alpha$ forest corresponds to the large number of
absorption features located blueward of the Lyman $\alpha$ emission
line in the spectra of quasi-stellar objects (QSOs).  In the last
decade, thanks to the power of 10-m class telescopes equipped with 
high-resolution spectrographs,
considerable progress has been made toward understanding the nature of
these absorption features.  This population of discrete lines is due
to HI absorption arising from the filamentary structure of the
cosmic-web, and in the approximation of a Gaussian velocity dispersion
can be well described by a series of Voigt profiles (Rauch 1998). A
variable amount of absorption in the \lya forest is also due to
ultraviolet (UV) transitions from heavy element ions which are usually
associated with strong Lyman $\alpha$ lines.

Quantities often studied in analyses of QSO spectra include the power
spectrum of the flux distribution, which is used to constrain
cosmological parameters and probe the dark matter power spectrum on
scales of order of 50 $h^{-1}$ Mpc (e.g. Croft et al. 2002, Viel et
al. 2004, McDonald et al. 2006).  Another quantity thoroughly studied
in the last few years is the flux probability distribution function
(PDF), which is sensitive not only to the spatial distribution of dark
matter, but also to the thermal state of the intergalactic medium
(IGM).  Previous studies of the Lyman $\alpha$ forest PDF have shed
light on the physical state of the IGM mostly at $z<3$.  Bolton et
al. (2008) compared the flux PDF measured from a sample of QSO spectra
presented by Kim et al. (2007, hereafter K07) to a set of
hydrodynamical simulations of the Lyman $\alpha$ forest, exploring
different cosmological parameters and various thermal histories.
Agreement between the data and simulations was obtained by adopting a
power-law temperature-density relation, $T=T_0 \Delta^{\gamma-1}$ (Hui
\& Gnedin 1997), where the low density IGM ($\Delta = \rho /
\langle\rho\rangle \leq 10$) was close to isothermal (i.e. with
$\gamma\sim 1$) or inverted ($\gamma < 1$). This result implied that
the low-density regions of the IGM may be considerably hotter and
their thermal state more complicated than usually assumed at these
redshifts.

Radiative transfer effects during the epoch of HeII reionization,
which should occur at $z>3$, are expected to play a non negligible role 
in setting this 
temperature-density relation.  Indeed, evidence that the tail-end of HeII
reionization is occurring at $z\simeq 3$ is supported by a variety of 
observations, such as the opacity of the HeII \lya forest (Reimers et
al. 1997; Heap et al. 2000; Kriss et al. 2001, Shull et al. 2010,
Worseck et al. 2011, Syphers et al. 2011), an increase in the
temperature of the IGM at mean density toward $z\simeq 3$ (Schaye et
al. 2000, Theuns et al. 2002, Lidz et al. 2010, Becker et al. 2011,
but see McDonald et al. 2001, Zaldarriaga, Hui, \& Tegmark 2001), the
observed evolution of the CIV/SiIV metal-line ratio (Songaila \& Cowie
1996; Songaila 1998, but see Kim, Cristiani \& D'Odorico 2002) and the
evolution of the \lya forest effective optical depth (Bernardi et
al. 2003, Faucher-Giguere et al. 2008, P{\^a}ris et al. 2011, 
but see Bolton, Oh \& Furlanetto 2009b).  

However, recent work suggests that photo-heating during HeII
reionization alone is not sufficient to achieve an inverted
temperature-density relation, $\gamma<1$ (McQuinn et al. 2009; Bolton,
Oh \& Furlanetto 2009a).  A recently proposed alternative channel for
heating, which naturally produces an inverted temperature-density
relation, may be provided by very energetic gamma rays from blazars
(Chang et al 2011, Puchwein et al. 2011).  In this scenario, very high
energy gamma rays annihilate and pair produce on the extragalactic
background light.  Powerful plasma instabilities then dissipate the
kinetic energy of the electron-positron pairs locally, volumetrically
heating the IGM.  On the other hand, continuum placement errors could
partly explain the preference of the K07 data for an inverted
temperature-density relation (see Lee 2012 for a recent discussion).
Indeed, Bolton et al. (2008) and more recently Viel et al. (2009)
found an isothermal temperature-density relation was consistent
(within $1.5\sigma$) with the K07 PDF constraints when treating
continuum uncertainties conservatively.

Ultimately, however, for further progress in understanding the origin
of an inverted temperature-density relation, additional observational
constraints on the IGM thermal state are required.  An investigation
of the flux probability distribution at redshifts around $z\sim 3$ or
higher may therefore shed more light on the possible inversion of the
temperature-density relation.  Such a measurement is also
complementary to other constraints on the IGM thermal state, such as
those presented recently by Becker et al. (2011), where the
temperature-density relation is not measured directly.  Such an
investigation is the main motivation of the present work.

In this paper, we use eight high resolution QSO spectra observed at
$3.3\le z \le 3.8 $ to extend the study of the Lyman $\alpha$ forest
PDF by K07 to higher redshifts.  The aim of this paper is
two-fold. First, we perform a detailed study of the redshift evolution
of the flux PDF across a wide redshift range.  This enables us to
study possible systematic trends in the PDF due to redshift-dependent
effects, such as the presence of metal absorption lines. In order to
corroborate our analysis, we also measure the Lyman $\alpha$ effective
optical depth and compare it to different estimates by other
authors. In the second step, the measured flux PDF is compared to
theoretical results obtained from cosmological hydrodynamical
simulations.  This enables us to investigate the thermal state of the
IGM during this interesting redshift range.

This paper is organized as follows. In Section 2, we present the QSO
spectra and the data analysis, and in Section 3 we present our
measurements of the flux PDF and effective optical depth. In Section
4, we introduce the hydrodynamical simulations used to obtain
constraints on the thermal state of the IGM from the PDF and, in
Section 5, we report the results from the analysis of the simulations.
Finally, in Section 6, we present our conclusions.

\section{The data set} 

The data set used in this paper consists of eight high-redshift QSO
spectra.  Two of the spectra (PKS2126-158 and PKS2000-330) were
obtained with the Ultraviolet and Visual Echelle Spectrograph (UVES)
(Dekker et al. 2000) at the Kueyen unit of the European Southern
Observatory (ESO) VLT (Cerro Paranal, Chile) as part of the ESO Large
Programme (LP): `The Cosmic Evolution of the IGM' (Bergeron et
al. 2004).  PKS1937-101 is from the program Carswell et
al. 077.A-0166(A), Q1317-0507, SDSSJ16212-0042 and Q1249-0159 are from
Kim et al. 075.A-0464(A), Q1209+0919 is from the program
Dessauges-Zavadsky et al. 073.B-0787(A) and Q0055-269 is from
D'Odorico et al. 65.O-0296(A).  All the spectra have been reduced with
the standard UVES pipeline provided by ESO.  Wavelengths have been
corrected to vacuum-heliocentric.

\begin{table*}
\vspace{0cm}
\caption[]{The QSO sample used in this work.  In the first column 
  the name of the QSO is reported.  Subsequent columns list the
  emission redshift of the QSO, the redshift range covered by the
  Lyman $\alpha$ forest, the wavelength range covered by the Lyman
  $\alpha$ forest and the signal to noise (S/N, see text for details).
  Note the wavelength range of the Lyman $\alpha$ forest excludes the
  region 4000 km s$^{-1}$ bluewards of the Lyman $\alpha$ emission
  line in order to avoid the proximity effect (see Sect.\ref{voigt}).}
%
%
\begin{tabular}{l|l|l|l|l|l}
\noalign{\smallskip} \hline \hline \noalign{\smallskip} QSO &
$z_{\rm em}$$^a$ & $z_{\rm Ly\alpha}$ & $\lambda_{\rm Ly\alpha}$ (\AA)& S/N &
Notes\\ \noalign{\smallskip} \hline \noalign{\smallskip} PKS2126-158 &
3.292 & 2.68-3.23 & 4473-5148 & 110 & 1 LLS ($z$=2.967), \\ & & & & &
2 Sub-DLAs ($z$=2.638, 2.769) \\ Q1209+0919 & 3.295 & 2.62-3.23 &
4404-5150 & 25 & 8 LLS \\ Q0055-269 & 3.66 & 2.93-3.59 & 4775-5583 &
51 & \\ Q1249-0159 & 3.66 & 2.94-3.60 & 4784-5594 & 70 & 3 LLS
($z$=3.102, 3.5243, 3.5428) \\ SDSSJ16212-0042 & 3.70 & 2.97-3.63 &
4821-5637 & 93 & 2 LLS ($z$=3.106, 3.144) \\ Q1317-0507 & 3.71 &
2.98-3.66 & 4840-5659 & 86 & 1 LLS ($z$=3.288) \\ PKS2000-330 & 3.78 &
3.03-3.72 & 4903-5733 & 68 & 2 LLS ($z$=3.189, 3.33) \\ PKS1937-101 &
3.79 & 3.04-3.72 & 4910-5742 & 150 & \\ \hline
\hline                   
\end{tabular}
\begin{flushleft}
$^a$ In some cases, the emission redshift $z_{\rm em}$ estimated on the
  basis of the QSO Lyman $\alpha$ emission line is an underestimation
  of the true value (see K07). If Lyman $\alpha$ absorption lines with
  $z>z_{\rm em}$ are present, we assume the highest redshift of the
  absorption line as a proxy for $z_{\rm em}$.
\end{flushleft}
\label{tab1}
\end{table*}


\begin{figure*}
\centering
\vspace{0.001cm}
\epsfig{file=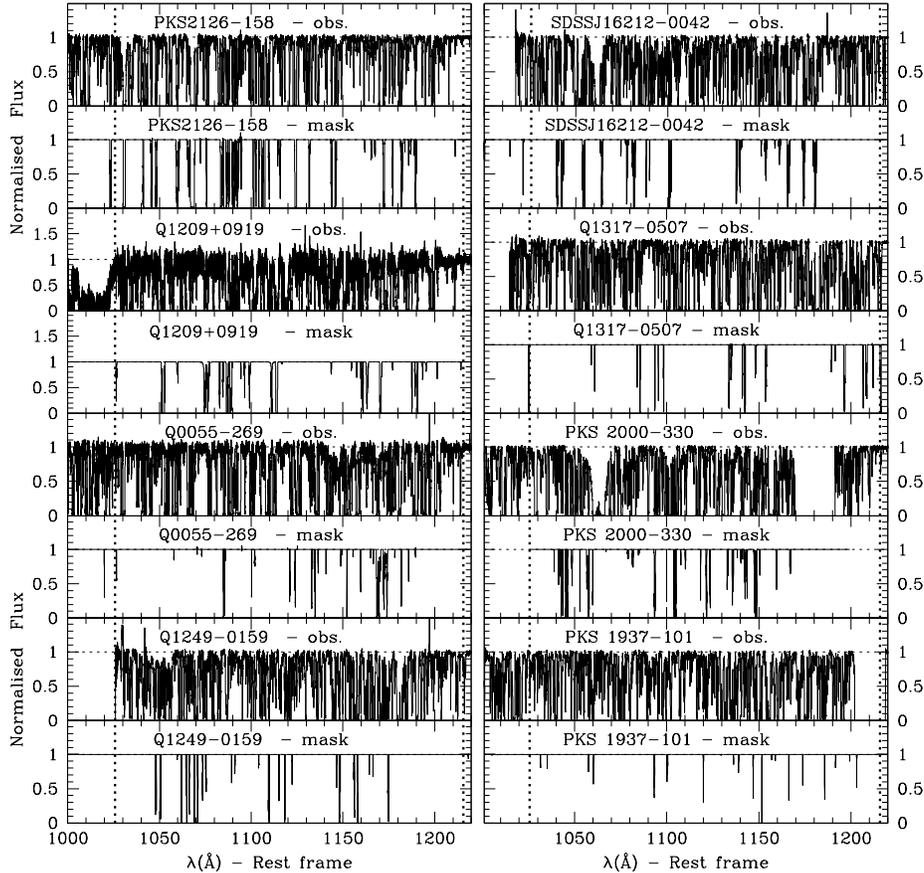,height=13cm,width=13cm}
\vspace{-0.5cm}
\caption{The eight QSO spectra which comprise our sample.  Each
  spectrum is normalised to the quasar continuum and the wavelengths
  are displayed in the QSO rest-frame.  For each QSO, the upper panel
  shows the normalised spectrum, whereas the lower panel displays all
  the spectral regions containing metal lines.  To measure the PDF,
  these regions have been masked following the criteria described in
  Sect.~\ref{mask}.  In each panel, the left dotted line and the right
  dotted line indicate the wavelengths of the Lyman $\beta$ and Lyman
  $\alpha$ transitions. The regions with no flux, such as the one
  apparent in the spectrum of PKS 2000-330 centered at $\lambda \sim
  1180$ \AA, are due to the wavelength gaps caused by using three
  separate CCDs in the dichroic setting.}
\end{figure*}

The properties of the QSO spectra used in this paper are summarized in
Table 1. The velocity resolution of each spectrum is 6.7 km s$^{-1}$
full-width half maximum, and each spectrum is binned in 0.05 \AA~
pixels. The signal-to-noise (S/N) ratio has been calculated by
considering regions of the spectra redwards of the QSO emission peak,
and corresponds to the inverse of the standard deviation of the
average normalized flux for pixels free of absorption. For some QSOs,
the spectra have already been used by D'Odorico et al. (2010) for the
study of the evolution of CIV in the IGM.  The spectra of three QSOs
(Q1317-0507, Q1249-0159 and SDSSJ16212-0042) have been reduced for
this work.

Most of the QSOs include one or more Lyman Limit systems (LLS),
i.e. absorption systems with neutral column density $17.2 \le
\log(N_{\rm HI}/{\rm cm}^{-2}) \le 19$.  For the comparison of our
data to simulation results, we have therefore excised all the pixels
belonging to the spectral regions included in the range $\lambda_{\rm
  LLS}-50$ \AA~ $\le \lambda \le \lambda_{\rm LLS}+50$ \AA, where
$\lambda_{\rm LLS}$ is the central wavelength of the LLS. PKS2126-158
additionally includes two sub-damped Lyman $\alpha$ systems ($19 \le
\log(N_{\rm HI}/{\rm cm}^{-2}) \le 20.3$).  The spectral regions where
these systems are present have also been excised from the spectrum,
following the same criterion as for LLS.  In addition, the
  identified metal lines associated with both LLS and sub-DLAs have
  been removed from the spectra as described in Sect.~\ref{mask}.

The two QSOs PKS2126−158 and
  Q0055-269 were also included in the K07 sample. 
  It is worth noting that the wavelength ranges for PKS2126−158 and
  Q0055-269 used in this work are different to those reported in
  Tab. 1 of K07.  In the case of PKS2126−158, our blue wavelength
  cut is different to the K07 value due to the presence of two
  sub-DLAs at $z=2.638$ and $z=2.769$.  We excluded the two regions
  contaminated by the sub-DLAs as described above, and a small region
  between the two sub-DLAs was left in our spectrum, extending from
  4470 \AA \, to 4540 \AA.  K07 instead decided to remove this region
  and cut their spectrum at 4630 \AA.  The red wavelength cut was
  chosen by K07 in order to avoid having pixels at wavelengths $>5112$
  \AA \, corresponding to redshifts $z>3.2$.  However, this redshift
  range is important for studying the possible effects of HeII
  reionization and so we include it in this work.  In the case of
  Q0055-269, the blue cut was performed in our spectrum at a
  wavelength corresponding to the Lyman $\beta$ emission line, whereas
  in the case of K07 they performed it at $4785$ \AA \, in order to
  exclude a metal absorption system at 4784 \AA. The red cut was
  performed by K07 at $5112$ \AA \, to again discard pixels with
  $z>3.2$.

In Fig. 1, we show the reduced spectra for all the QSOs of our sample
and, for each spectrum, the contamination from  metal lines. The
regions containing metal lines have been masked following the criteria
described in Sect.~\ref{mask}. In Fig. 2, for each QSO spectrum, the
redshift coverage and the Lyman $\alpha$ forest wavelength coverage
are shown.

\begin{figure}
\centering
\vspace{0.001cm}
\epsfig{file=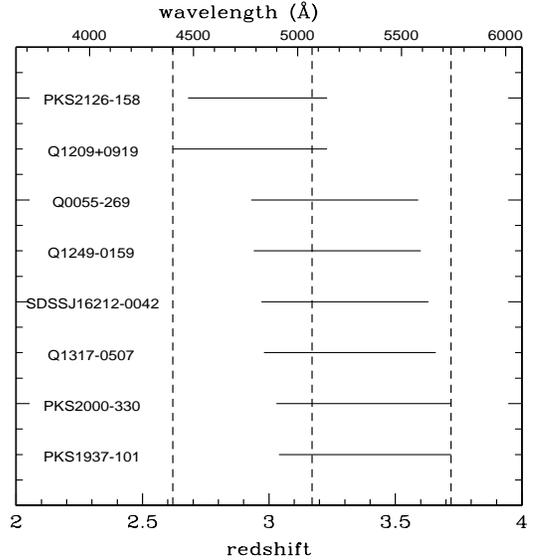,height=8cm,width=9cm}
\caption{The solid lines display the redshift (and wavelength, see
  upper axis) coverage of the \lya forest in the QSO spectra used in
  our sample. The dashed lines at $z=2.62$, $z=3.17$ and $z=3.72$
  represent the extremes of the two redshift bins used to calculate
  the flux probability distribution function, as described in
  Sect.~\ref{PDF}. }
\label{zbin}
\end{figure}

\subsection{Continuum fitting}

In each spectrum all the absorption lines redwards of the Lyman
$\alpha$ emission line, as well as those belonging to the Lyman
$\alpha$ forest, have been fitted with Voigt profiles.  To
satisfactorily perform this task, for each spectrum a precise estimate
of the continuum is required.  However, the problem of continuum
fitting high resolution data requires particular attention, and it has
been widely discussed in the literature (e.g. K07; Desjacques et
al. 2007; Faucher-Giguere et al. 2008).   An automatic procedure
  to fit QSO spectra has also been developed by Dall'Aglio et
  al. (2008a; 2008b).  They used an extended sample of high
  resolution, high S/N QSO spectra, and by means of an adaptive
  techinque involving cubic spline interpolations and a Monte-Carlo
  method, they were able to estimate the QSO continua and relative
  uncertainties.  In this paper, however, continuum fitting was
  performed on our smaller data set as described below.  It will be
  interesting to use a larger sample in the future to develop
  automatic techniques for continuum fitting and investigate
  associated uncertainties further.

\begin{figure*}
\centering
\vspace{0.001cm}
\epsfig{file=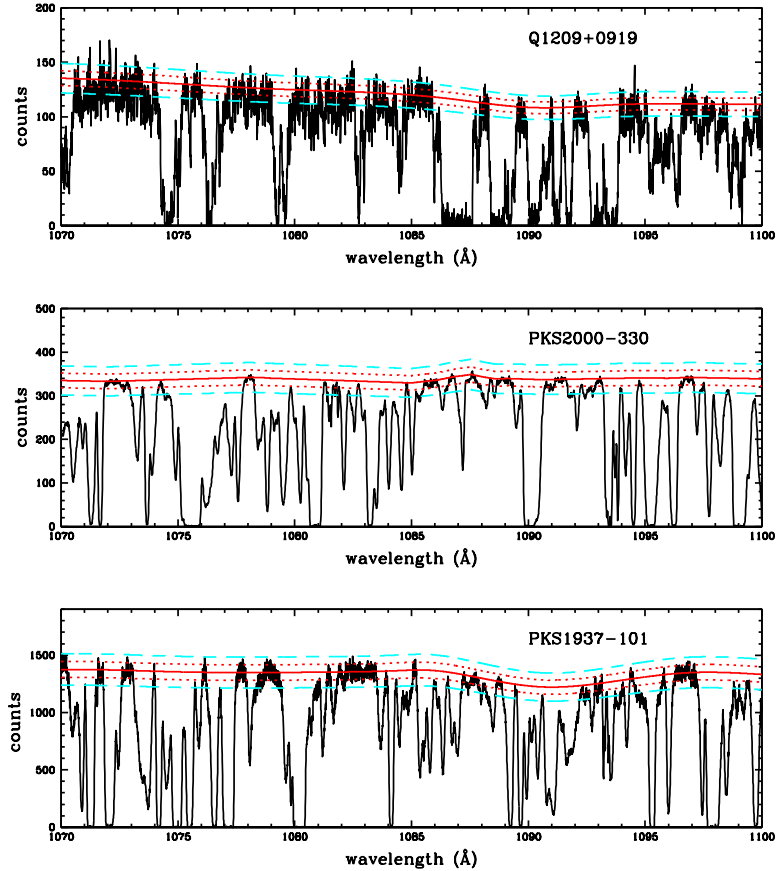,height=13cm,width=13cm}
\caption{An illustration of the uncertainty on the continuum placement
  for three of the spectra in our sample with different S/N values:
  S/N=25 for Q1209+0919 (upper panel), S/N=68 for PKS2000-330 (middle
  panel) and S/N=150 for PKS1937-101 (lower panel).  In each panel,
  the final continuum fit is shown by the solid curve, while the
  dotted and dashed curves show the fitted continuum after it is
  increased/decreased by 5 per cent and by 10 per cent, respectively. }
\label{cont}
\end{figure*}

For each QSO, the continuum was fitted using the IRAF package, by
means of high-order polynomial functions.  The continuum fitting was
performed by cutting each spectrum into several chunks, depending on
the S/N and on the number of absorption lines, and then joining the
chunks back together into one single spectrum. The continua were
determined by using regions free of absorption lines.  This task does
not present substantial difficulties when the region redwards of the
QSO Lyman $\alpha$ emission line is considered, since this region
presents only a few (metal) absorption lines.  However, regions
bluewards of the QSO Lyman $\alpha$ emission lines, such as the Lyman
$\alpha$ forest, are instead populated by a large number of absorption
lines. In general, the higher the emission redshift of the QSO, the
greater the average Lyman $\alpha$ forest opacity. This makes
continuum fitting toward higher redshift more difficult, since regions
which are free of absorption lines become increasingly small and
rare.  In this case, a linear interpolation between two subsequent
absorption-free chunks is used as a first-order guess for the
continuum level.
Another major problem in continuum fitting concerns the possible
presence of wide and shallow absorption features in the spectra.
Large-scale absorption can sometimes be incorrectly attributed to
unabsorbed parts of the spectrum.  This represents a problem in this
procedure, since it can lead to an underestimate of the true
continuum.  To avoid such mistakes, it is necessary not only to check
carefully every small chunk of the spectrum which is being fitted, but
also the joint spectrum by searching for possible wide depressions due
to large-scale absorption features which are not visible while working
on the individual chunks.  All the spectra and the fitted continua
have been repeatedly and independently checked in this manner in order
to identify such features and to avoid continuum underestimates.

As discussed in K07, the uncertainty on the continuum depends on the
S/N, which varies not only from spectrum to spectrum, but may even
vary within one single spectrum.  As a general rule, the higher the
S/N, the smaller the continuum uncertainty.  This is illustrated in
Fig.~\ref{cont}\footnote{ In Fig.~\ref{cont}, the spectrum of
  PKS1937-101 appears to be noisier than PKS2000-330, contrary to what
  is reported in Tab. 1.  Recall, however, that the S/N reported in
  Tab. 1 have been calculated by means of absorption-free regions
  redwards of the QSO emission peak, and that S/N may increase with
  increasing wavelengths owing to the increase in the efficiency of
  the UVES CCDs. As a check, we have therefore recalculated the S/N in
  the Lyman $\alpha$ forest for PKS1937-101 and PKS2000-330 by
  computing the average ratio between the flux and relative error
  $<F/\sigma_F>$, finding in both cases $<F/\sigma_F>=59$. However,
  the maximum S/N value found in the \lya forest of PKS1937-101 is
  $<F/\sigma_F>_{max} \simeq 100$, whereas for PKS2000-330 it is
  $<F/\sigma_F>_{max} \simeq 62$. } where we show three unnormalized
spectra with their final fitted continuum (solid lines), and with the
continuum increased and decreased by 5 per cent (dotted lines) and by
10 per cent (dashed lines).  The cases depicted in Fig.~\ref{cont}
represent the spectrum with the lowest S/N in Tab. 1 (Q1209+0919,
upper panel), one case with an intermediate S/N value (PKS2000-330,
middle panel) and the spectrum with the highest S/N (PKS1937-101,
lower panel).  Although it is very difficult to precisely quantify the
uncertainty for the continuum determination, Fig.~\ref{cont}
demonstrates that, in the case of Q1209+0919, an uncertainty of the
order of 10 per cent is exaggerated, but an uncertainty of 5 per cent
seems plausible.  As stressed earlier, this case represents the
spectrum with the worst S/N in our sample.  In the other two cases, it
appears that a continuum which is decreased or increased by 5 per cent
represents an underestimation and overestimation, respectively, of the
actual continuum. For this reason, a typical uncertainty of 5 per cent
should be regarded as conservative and representative of most of the
spectra in our sample, in agreement with the analysis of K07 and the
one of Dall'Aglio et al. (2008a). Later on, the effects of the
continuum estimate on the flux PDF will be investigated in more
detail.

\subsection{Voigt Profile fitting}
\label{voigt}

Once the continuum fitting was completed, in each spectrum, all the
absorption features blueward of the QSO Lyman $\alpha$ emission line
were then fitted with Voigt profiles.  The aim of this procedure is to
determine, for each absorption line, the redshift $z$, the Doppler
parameter $b$ and the column density $N$ of the physical system giving
rise to that absorption feature.

The Lyman $\alpha$ forest includes the spectral region between the
wavelengths of the Lyman $\beta$ emission line,
$\lambda_{\beta}=1025.72 \, (1+z_{\rm QSO})$, and the wavelength
corresponding to the Lyman $\alpha$ emission line,
i.e. $\lambda_{\alpha}=1215.67 \, (1+z_{\rm QSO})$.  To avoid
contamination from the QSO line-of-sight proximity effect, in each
spectrum we excluded an interval of 4000 km s$^{-1}$ bluewards of the
Lyman $\alpha$ emission line. All the absorption features were then
identified and fitted by means of the
RDGEN\footnote{http://www.ast.cam.ac.uk/~rfc/rdgen.html},
VPFIT\footnote{http://www.ast.cam.ac.uk/~rfc/vpfit.html} and
VPGUESS\footnote{http://www.eso.org/~jliske/vpguess/} packages.

Firstly, all the metal absorption lines redwards of the QSO Lyman
$\alpha$ emission peak were identified and fitted. This task does not
present significant difficulties, since the number of absorption lines
in this spectral region is relatively low and lines due to different
transitions are rarely blended.  Secondly, the identification of heavy
element absorption in the Lyman $\alpha$ forest was performed.  This
task requires more attention owing to the large amount \lya lines in
this region and to the frequent presence of broad, saturated
absorption features.  In this spectral region, isolated metal lines
are very rare. Most of the metal absorption lines are blended with
Lyman $\alpha$ lines, complicating the precise identification of metal
line profiles.  A common technique to search for metal absorption in
the Lyman $\alpha$ forest therefore employs the use of velocity width
profiles.  Once a strong transition is identified at wavelengths
redward of the QSO Lyman $\alpha$ emission line, e.g. due to CIV or
SiIV, associated ionic transitions such as SiII, SiIII, CII at the
corresponding redshift and velocity profile can be searched for in the
Lyman $\alpha$ forest.

\begin{figure}
\centering
\vspace{0.001cm}
\epsfig{file=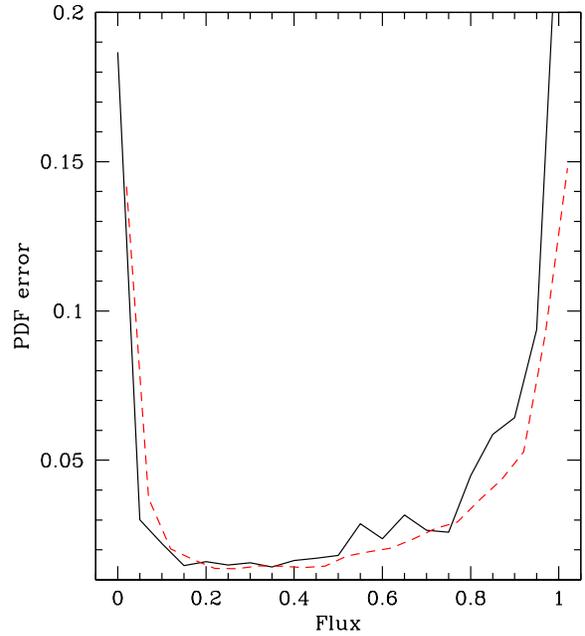,height=9cm,width=9cm}
\caption{The flux PDF errors as a function of the normalised flux
  measured from all eight QSOs, computed by means of the jackknife
  (black solid lines) and the bootstrap (red dashed line) methods.}
\label{fig3}
\end{figure}

When metal lines in the Lyman $\alpha$ forest are heavily blended with
HI absorption, it is generally difficult to identify the shape of the
profile.  The parameters which are most difficult to recover in this
case are the Doppler parameter $b$ and the column density $N$.  In
these cases, as an initial guess for VPFIT we provide the profile of
the lines associated with the systems already identified redwards of
the QSO emission, usually satisfactorily fitted by VPFIT.  VPFIT then
corrects the initial guess and modifies it in order to obtain an
improved fit to that spectral region. Once a stable solution is
obtained which is not sensitive to the initial guess and characterised
by a satisfactory reduced $\chi^{2}$ value (typically of the order of
$1-3$, depending on the noise level of the spectrum) the fit process
is complete.  An alternative way to identify metal absorption in the
forest is based on the value of the Doppler parameter, which is in
general narrower than for Lyman $\alpha$ lines (e.g. Tescari et
al. 2011). However, in this analysis any absorption not corresponding
to an identified heavy element transition has been assumed to be due
to Lyman $\alpha$ absorption.

\subsection{Metal removal from the spectra}
\label{mask}

The presence of metal absorption alters the transmitted fraction in
the QSO spectra and consequently the shape of the PDF.  For this
reason, in order to use the PDF as a probe of the physical state of
the Lyman $\alpha$ forest, it is necessary to remove the metal
absorption from the \lya forest in the QSO spectra.

Once the metal identification and Voigt profile fitting is complete,
for each QSO we remove metal lines from the spectra in the following
manner.  The pixels belonging to spectral regions contaminated by
identified metal absorption are excised from the spectra using a
quantitative criterion which depends on the Doppler parameter of the
metal absorption line, $b$.  All the pixels at wavelengths, $\lambda$,
within the interval $\lambda_0 - \Delta \lambda \le \lambda \le
\lambda_0 + \Delta \lambda$ were removed from the spectra, where
$\lambda_0$ is the central wavelength of the metal absorption line and
$ \Delta \lambda=\lambda_{Z,0} \, (1+z')$.  Here $\lambda_{Z,0}$ is
the rest-frame wavelength of the transition of a given element $Z$ and
$z'=z-dz$, where $z$ is the redshift of the absorbing system and
$dz=\frac{2 \, b \sqrt{\ln\, 2} \, (1+z)}{c}$.  For each QSO spectra,
we created a ``mask'' using this procedure, which is then used to
remove the pixels belonging to the metal-contaminated regions.

The completeness of the masking is dependent on the success of
identifying metals. It is clearly impossible to identify all the metal
absorption in the Lyman $\alpha$ forest.  In this work we are dealing
with QSO spectra at $z>3$ which contain a large number of HI
absorption lines.  At these redshifts, metal lines are frequently
blended with strongly saturated HI absorption, and when this is the
case their contribution to the total absorption is negligible.
However, there will be very few cases where the metal lines are
isolated and not identified.  For this reason, we believe that the
results presented in this paper are not substantially affected by
incomplete metal identification in the Lyman $\alpha$ forest.  This
issue will be discussed further in the section dedicated to the
results.

\begin{figure*}
\centering
\vspace{0.001cm}
\epsfig{file=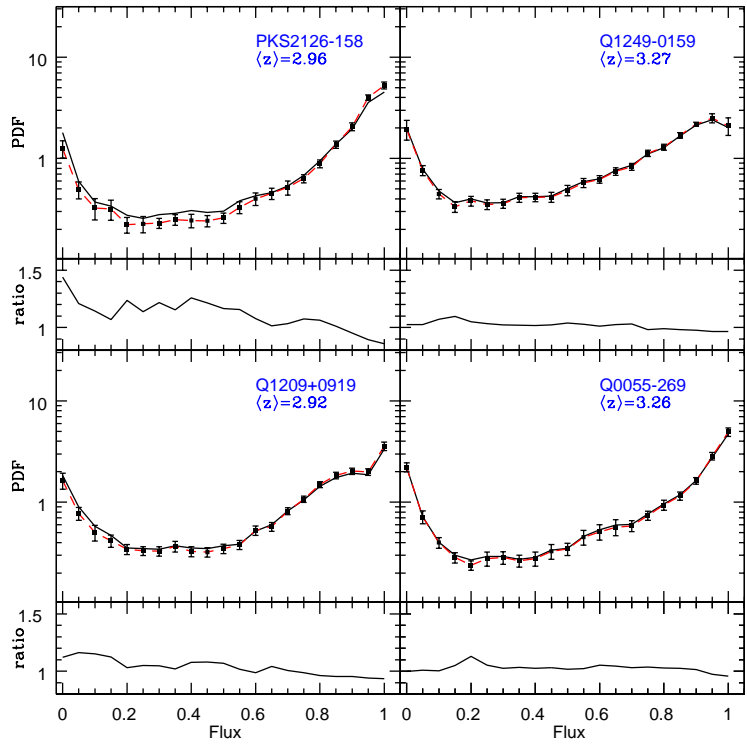,height=10cm,width=10cm}
\caption{The effect of metal line absorption on the PDFs obtained for
  four of the QSOs in our sample.  In the main panels, the solid
  curves display the PDFs calculated for each spectrum before removing
  metal lines.  The solid squares with error bars show the PDFs
  calculated for the masked spectra, i.e. after removing the metal
  absorption. The dashed lines are drawn to guide the eye through the
  solid squares. For each QSO, the name and the average redshift of
  Lyman $\alpha$ forest absorption is indicated.  In the smaller
  panels, the solid curve represents the ratio between the PDF
  calculated before and after removing metal lines.  }
\label{fig4}
\end{figure*}

\begin{figure*}
\centering
\vspace{0.001cm}
\epsfig{file=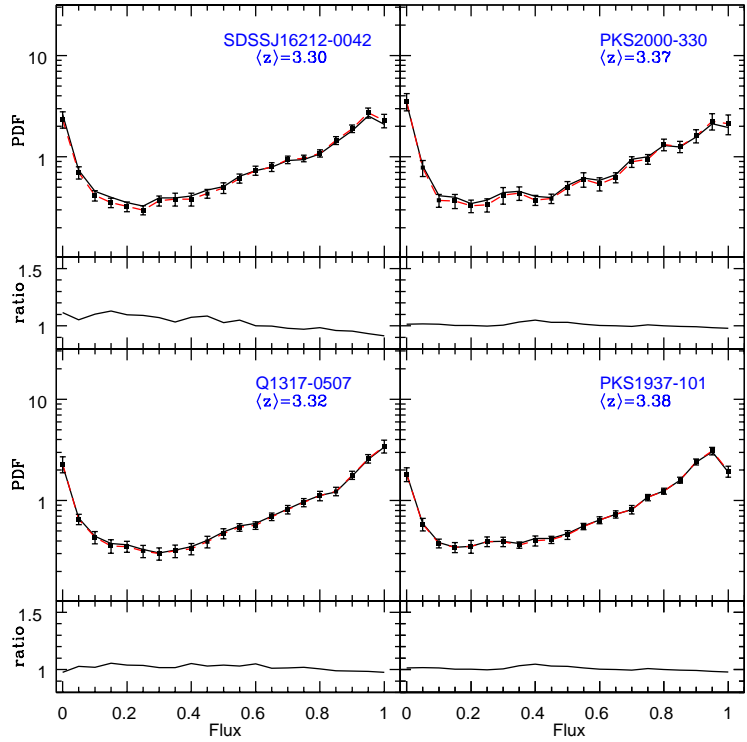,height=10cm,width=10cm}
\caption{As in Fig.\ref{fig4}, but now for the other four QSOs in
  our sample. }
\label{fig5}
\end{figure*}

\section{Data Analysis and results} 
\label{PDF}

Once each spectrum has been normalised to the continuum, i.e.  once
the quantity $F=S/C$ has been calculated for each pixel, where $S$ and
$C$ are the observed flux and the estimated continuum, respectively,
the flux probability distribution function can be measured.  For each
flux bin $F$, the PDF is calculated from the ratio between the number,
$N_{\rm F,F+\Delta F}$, of pixels with flux between $F$ and $F+\Delta
F$ and the total number, $N_{\rm tot}$, of pixels in the spectral
region corresponding to the Lyman $\alpha$ forest (K07):

\begin{equation}
{\rm PDF(F)} =\frac{N_{\rm F,F+\Delta F}}{N_{\rm tot}}.  
\end{equation}
The PDF is measured using this procedure in bins of width $\Delta
F=0.05$ (McDonald et al. 2000; K07). We have assigned flux values
$F=0$ and $F=1$ to all the pixels characterised by flux levels
$F<0.025$ and $F>1$, respectively.  The PDF is calculated both for
single QSO spectra, and for the full sample. To study the redshift
evolution of the PDF, we also divide the full sample in two redshift
bins. The redshift bin width is chosen in order to have comparable
numbers of pixels in each bin.

\subsection{Error bar estimates}
\label{errest}
The error bars are estimated by means of a jackknife method, as
described by Lidz et al. (2006).  Once the PDF for a single QSO or for
the full sample in a given redshift bin has been calculated, the
contributing spectra are then divided into $n_{\rm c}$ chunks of width
$\sim 50$ \AA.  If the PDF in the flux bin $F_{\rm i}$ is
$\widehat{P}(F_{\rm i})$ and the PDF estimated without the $k$-th
chunk in the flux bin $F_{\rm i}$ is $\widetilde{P}_{\rm k}(F_{\rm
  i})$, the covariance matrix cov$(i,j)$ between the PDF in flux bin
$F_{\rm i}$ and the PDF in flux bin $F_{\rm j}$ can be defined as
\begin{equation}
\mathrm{cov}(i,j) = \sum_{k=1}^{n_{c}} [\widehat{P}(F_i)-\widetilde{P}_{k}(F_i)]
[\widehat{P}(F_j)-\widetilde{P}_{k}(F_j)],
\end{equation}
For the flux bin $F_{\rm i}$, the errors computed with the jackknife method,
$\sigma_{\rm J}$, are given by the square root of the diagonal of the matrix
$\rm cov(i,j)$, i.e.
\begin{equation}
\sigma_J=\sqrt{\mathrm{cov(i,i)}}.   
\end{equation}
In addition, it is useful to have an alternative way to estimate the
error bars for comparison to the uncertainties calculated using the
jackknife method.  We therefore also estimate error bars with a
bootstrap technique.  This technique consists of dividing the spectra
into chunks of $\sim 5$ \AA$\,$, and drawing a large number (500) of
these chunks from the sample randomly with replacement.  The PDF
averaged over the 500 realizations coincides with the PDF of the
random sample.  For each flux bin, the uncertainty is then estimated
by calculating the standard deviation of the PDFs obtained by sampling
the 500 chunks with replacement.

In figure \ref{fig3}, the 1$\sigma$ errors for the PDF, estimated
using the jackknife and bootstrap techniques, are compared.  The PDF
is computed from the full sample in this instance, without dividing
the spectra into different redshift bins.  Fig. ~\ref{fig3} shows
  how the two methods provide comparable error estimates, although the
  jackknife estimates are larger at $F=0$ and $F=1$.  In the
remainder of the paper, unless otherwise stated, the uncertainties for
the PDF will be computed by using the jackknife method.

\begin{table*}
\caption{The mean PDF measured from the full sample, divided into two
  redshift bins and calculated by (1) using all pixels, (2) after
  removing contamination from metal absorption lines and (3) after
  removing Lyman limit systems and metal lines.  Note that the
  redshift bins in the final two columns have a slightly different
  mean redshift due to the removal of LLS.  The redshift bins have
  been slightly modified in order to have comparable numbers of pixels
  in each bin.}
\begin{tabular}{lcccccccc}
\hline
  &   All pixels &   & No metals  &  &  No metals, No LLS &  &  & \\  
  &    (1) &   & (2)  &  &  (3) &  &  & \\  
\hline
\noalign{\smallskip}
$F$  & $<z> \, =2.90$  &  $<z> \, =3.45$  & $<z> \, =2.90$  &  $<z> \, =3.45$  &  $<z> \, =2.95$  &  $<z> \, =3.48$  \\ 
\noalign{\smallskip}
\hline
\noalign{\smallskip}
  0.00  &    1.76 $\pm$   0.18  &    2.37  $\pm$  0.24    &        2.00 $\pm$   0.16  &    2.38 $\pm$   0.23  &   1.54  $\pm$  0.11   &   2.28 $\pm$   0.30   \\ 
  0.05  &    0.66 $\pm$   0.05  &    0.71  $\pm$  0.04    &        0.74 $\pm$   0.04  &    0.72 $\pm$   0.04  &   0.57  $\pm$  0.04   &   0.70 $\pm$   0.06   \\ 
  0.10  &    0.41 $\pm$   0.03  &    0.41  $\pm$  0.03    &        0.46 $\pm$   0.03  &    0.43 $\pm$   0.03  &   0.34  $\pm$  0.03   &   0.42 $\pm$   0.04   \\ 
  0.15  &    0.36 $\pm$   0.02  &    0.34  $\pm$  0.02    &        0.39 $\pm$   0.02  &    0.36 $\pm$   0.02  &   0.30  $\pm$  0.03   &   0.35 $\pm$   0.03   \\ 
  0.20  &    0.30 $\pm$   0.01  &    0.33  $\pm$  0.02    &        0.32 $\pm$   0.01  &    0.35 $\pm$   0.02  &   0.26  $\pm$  0.02   &   0.35 $\pm$   0.02   \\ 
  0.25  &    0.30 $\pm$   0.02  &    0.33  $\pm$  0.02    &        0.32 $\pm$   0.02  &    0.34 $\pm$   0.02  &   0.26  $\pm$  0.03   &   0.35 $\pm$   0.03   \\ 
  0.30  &    0.30 $\pm$   0.02  &    0.36  $\pm$  0.02    &        0.32 $\pm$   0.02  &    0.37 $\pm$   0.02  &   0.27  $\pm$  0.02   &   0.36 $\pm$   0.03   \\ 
  0.35  &    0.33 $\pm$   0.02  &    0.35  $\pm$  0.02    &        0.35 $\pm$   0.02  &    0.36 $\pm$   0.02  &   0.30  $\pm$  0.02   &   0.35 $\pm$   0.02   \\ 
  0.40  &    0.30 $\pm$   0.02  &    0.37  $\pm$  0.02    &        0.34 $\pm$   0.02  &    0.38 $\pm$   0.02  &   0.29  $\pm$  0.03   &   0.36 $\pm$   0.03   \\ 
  0.45  &    0.31 $\pm$   0.01  &    0.41  $\pm$  0.01    &        0.34 $\pm$   0.01  &    0.42 $\pm$   0.02  &   0.31  $\pm$  0.02   &   0.42 $\pm$   0.02   \\ 
  0.50  &    0.35 $\pm$   0.02  &    0.46  $\pm$  0.01    &        0.38 $\pm$   0.02  &    0.47 $\pm$   0.02  &   0.34  $\pm$  0.02   &   0.48 $\pm$   0.02   \\ 
  0.55  &    0.40 $\pm$   0.02  &    0.58  $\pm$  0.02    &        0.43 $\pm$   0.02  &    0.58 $\pm$   0.02  &   0.40  $\pm$  0.04   &   0.60 $\pm$   0.03   \\ 
  0.60  &    0.51 $\pm$   0.02  &    0.61  $\pm$  0.03    &        0.52 $\pm$   0.02  &    0.62 $\pm$   0.03  &   0.50  $\pm$  0.03   &   0.62 $\pm$   0.04   \\ 
  0.65  &    0.56 $\pm$   0.02  &    0.71  $\pm$  0.03    &        0.58 $\pm$   0.02  &    0.72 $\pm$   0.04  &   0.57  $\pm$  0.03   &   0.72 $\pm$   0.05   \\ 
  0.70  &    0.70 $\pm$   0.04  &    0.81  $\pm$  0.02    &        0.72 $\pm$   0.04  &    0.82 $\pm$   0.03  &   0.68  $\pm$  0.03   &   0.84 $\pm$   0.05   \\ 
  0.75  &    0.91 $\pm$   0.05  &    0.95  $\pm$  0.03    &        0.91 $\pm$   0.05  &    0.96 $\pm$   0.04  &   0.86  $\pm$  0.04   &   0.97 $\pm$   0.04   \\ 
  0.80  &    1.21 $\pm$   0.05  &    1.13  $\pm$  0.06    &        1.19 $\pm$   0.05  &    1.14 $\pm$   0.05  &   1.14  $\pm$  0.05   &   1.15 $\pm$   0.06   \\ 
  0.85  &    1.55 $\pm$   0.06  &    1.37  $\pm$  0.06    &        1.49 $\pm$   0.06  &    1.38 $\pm$   0.05  &   1.44  $\pm$  0.07   &   1.40 $\pm$   0.06   \\ 
  0.90  &    2.01 $\pm$   0.09  &    1.88  $\pm$  0.08    &        1.92 $\pm$   0.08  &    1.87 $\pm$   0.08  &   2.01  $\pm$  0.10   &   1.93 $\pm$   0.10   \\ 
  0.95  &    2.84 $\pm$   0.07  &    2.66  $\pm$  0.13    &        2.67 $\pm$   0.07  &    2.56 $\pm$   0.14  &   3.24  $\pm$  0.12   &   2.69 $\pm$   0.21   \\ 
  1.00  &    3.93 $\pm$   0.19  &    2.89  $\pm$  0.26    &        3.61 $\pm$   0.18  &    2.77 $\pm$   0.27  &   4.38  $\pm$  0.29   &   2.69 $\pm$   0.24   \\
\noalign{\smallskip}
\hline
\end{tabular}
\label{tab2}
\end{table*}

\begin{table*}
\caption{The HI effective optical depth and its dependence on the
  contamination of metal absorption lines, calculated after dividing
  each spectrum in two redshift bins.  In columns 1, 2 and 3 the name
  of the QSO, the Lyman $\alpha$ wavelength range and the average
  redshift are shown, respectively.  In columns 4, 5 and 6 we report
  the HI effective optical depth with its error, the optical depth
  computed without removing metal lines with its error, and the
  percentage contribution of metal absorption, respectively.}
\begin{tabular}{lccccccr}
\hline \noalign{\smallskip} QSO & wavelengths (\AA\/) & $<\!z\!>$ &
$\tau_{\rm eff}$ & $\tau_{\rm eff,HI+Z}$ & $\tau_{\rm eff,HI+Z}$/$\tau_{\rm eff}$  & metals (per cent) \\ \noalign{\smallskip}
\hline \noalign{\smallskip} 
PKS2126-158    &    	  4470-4880 &       2.85     &	      0.329  $\pm$    0.037  &    0.421  $\pm$    0.041  &      1.278  $\pm$ 0.190  & 28  \\         
      	       &    	  4880-5148 &  	    3.12     & 	      0.238  $\pm$    0.031  &    0.271  $\pm$    0.031  &      1.138  $\pm$ 0.197  & 14  \\  
Q1209+0919     &    	  4404-4780 &  	    2.78     &        0.373  $\pm$    0.031  &    0.423  $\pm$    0.037  &      1.132  $\pm$ 0.136  & 13  \\  
               &     	  4780-5150 &  	    3.08     &        0.425  $\pm$    0.040  &    0.450  $\pm$    0.039  &      1.060  $\pm$ 0.135  &  6  \\  
Q0055-269      &    	  4774-5170 &  	    3.09     &        0.339  $\pm$    0.033  &    0.344  $\pm$    0.032  &      1.014  $\pm$ 0.137  &  1  \\  
      	       &    	  5170-5583 &  	    3.42     &        0.429  $\pm$    0.036  &    0.439  $\pm$    0.035  &      1.024  $\pm$ 0.119  &  2  \\  
Q1249-0159     &           4784-5190 &      3.10     &        0.382  $\pm$    0.028  &    0.399  $\pm$    0.028  &      1.045  $\pm$ 0.106  &  4  \\  
      	       &           5190-5594 &      3.44     &        0.529  $\pm$    0.041  &    0.542  $\pm$    0.040  &      1.024  $\pm$ 0.110  &  2  \\  
SDSSJ16212-0042&           4821-5235 &      3.14     &        0.555  $\pm$    0.042  &    0.596  $\pm$    0.044  &      1.073  $\pm$ 0.113  &  7  \\  
               &    	  5230-5637 &  	    3.47     &        0.394  $\pm$    0.032  &    0.427  $\pm$    0.034  &      1.085  $\pm$ 0.123  &  8  \\  
Q1317-0507     &    	  4840-5240 &  	    3.15     &        0.384  $\pm$    0.031  &    0.388  $\pm$    0.031  &      1.010  $\pm$ 0.116  &  1  \\  
      	       &    	  5240-5659 &  	    3.48     &        0.495  $\pm$    0.038  &    0.499  $\pm$    0.039  &      1.008  $\pm$ 0.110  &  1  \\  
PKS2000-330    &    	  4903-5270 &  	    3.18     &        0.574  $\pm$    0.048  &    0.582  $\pm$    0.046  &      1.015  $\pm$ 0.117  &  1  \\  
      	       &    	  5270-5734 &  	    3.53     &        0.553  $\pm$    0.043  &    0.570  $\pm$    0.041  &      1.029  $\pm$ 0.108  &  3  \\  
PKS1937-101    &           4910-5323 &      3.21     &        0.371  $\pm$    0.027  &    0.376  $\pm$    0.027  &      1.014  $\pm$ 0.103  &  1  \\  
               &    	   5323-5742 &       3.55    &       0.487   $\pm$     0.037  &    0.492 $\pm$    0.036  &      1.011  $\pm$ 0.106   &  1  \\               
\noalign{\smallskip} \hline
\end{tabular}
\label{tab3}
\end{table*}

\subsection{The effect of the metals}
\label{metals}

The effect of metal absorption on the PDF is displayed in
Figures~\ref{fig4} and \ref{fig5}, where we show the PDF obtained
from each of the eight QSOs in our sample. In these two figures, the
solid and dashed curves in the large panels show the PDF computed
including and excluding metal absorption lines, respectively.  In the
small panels, for each QSO the ratio between the PDF calculated
including metals and after removing the metals is shown.  The average
Lyman $\alpha$ forest redshift spans from 2.96, for PKS2126-158, up to
3.38 in the case of PKS1937-101.

In general, the effect of removing metals on the PDF is to decrease
the number of pixels with low flux and increase the number of pixels
with high flux.  As a result, the PDF computed after metal removal is
slightly steeper than the PDF calculated before removing the metal
absorption.  This effect is most clearly visible for some of the QSOs
in Fig.~\ref{fig4}, such as PKS2126-158.  For this spectrum, the
effect of metal removal on the PDF is visible for flux values $F<0.6$,
 and is in agreement with the lower redshift analysis of K07, who find a PDF ratio 
similar to ours.

\begin{figure}
\centering
\vspace{0.001cm}
\epsfig{file=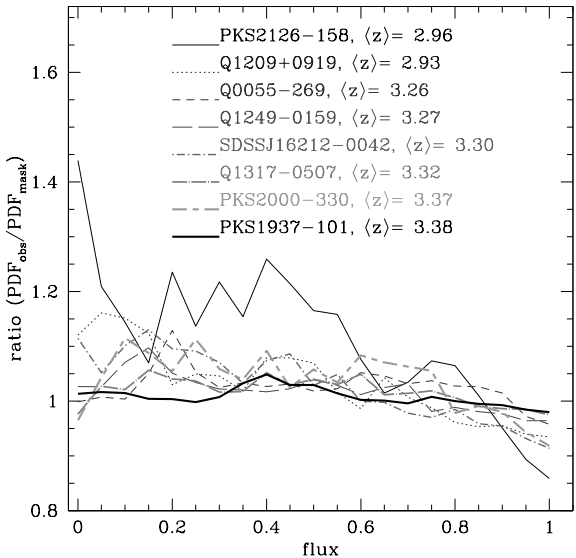,height=8cm,width=8cm}
\caption{The ratios between the PDFs measured for each of the eight
  QSO spectra analysed in this work before and after the removal of
  metal line absorption. Note that the line thickness increases with
  the average Lyman $\alpha$ forest redshift.}
\label{fig6}
\end{figure}

The effect of metal absorption on the PDF is also expected to depend on
the redshift of the QSO and hence the average redshift of the Lyman
$\alpha$ absorption.  The impact of metal absorption on the PDF is
expected to be more significant at lower redshifts, where Lyman
$\alpha$ forest opacity is generally lower (K07).  A weak trend with
redshift is indeed visible in Figs.~\ref{fig4}-\ref{fig5}, but it is
illustrated more clearly in Fig.~\ref{fig6}, where for each QSO we
show the ratio between the PDF calculated including metal lines and
the PDF computed after removing metals from the spectra.  The ratio is
higher than 1.2 for flux values $F<0.4$ and lower than 0.9 for $F>0.9$
only in one case, PKS2126-158, which is one of the QSOs with the
lowest value of $\langle z\rangle$ \footnote{For each QSO, $\langle
  z\rangle$ is the average redshift of the Lyman $\alpha$ forest range
  as reported in Tab. 1.}.  Q1209+0919 has a comparable $\langle
z\rangle$ and has a ratio close to 1.2 at $F<0.1$.  The other six QSOs
have similar $\langle z\rangle$ (between 3.26 and 3.38), and also
exhibit comparable levels of metal contamination. For these six QSOs
small fluctuations from one LOS to another are apparent, but the
ratios are all between 0.9 and 1.2 and are much smaller than the
variations for the QSOs at $\langle z\rangle < 3$.

The redshift dependence of metal absorption may be understood in terms
of the larger number of Lyman $\alpha$ absorption lines toward higher
redshift, which leave less room for isolated metal absorption
features.  This does not necessarily mean that QSOs at higher
redshifts \emph{have} less metal absorption systems along their LOS,
as the metal lines may instead be blended with Lyman $\alpha$
absorption lines.  At the same time, however, this observation does
not fully rule out the possibility the metal content of the
intergalactic medium is decreasing toward higher redshift.  We will
return to the issue of metal contamination of QSO spectra in
Sect.~\ref{tau}.

\begin{figure}
\centering
\vspace{0.001cm}
\epsfig{file=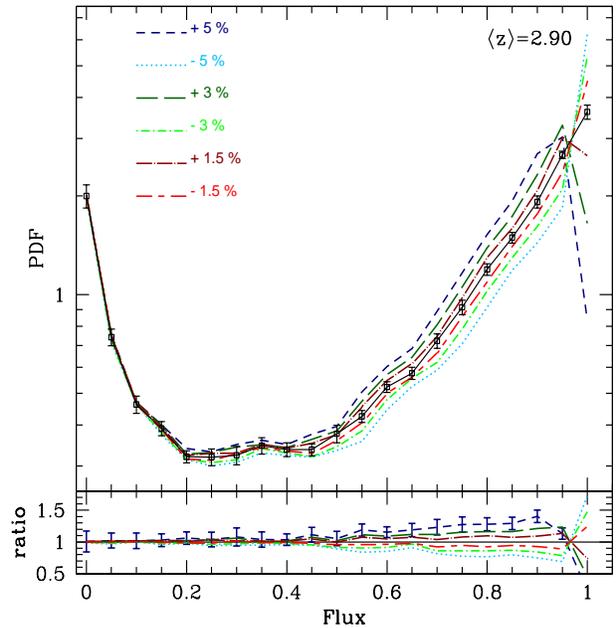,height=9cm,width=8.5cm}
\caption{The effect of continuum placement uncertainty on the PDF
  calculated for our full sample in the lower redshift bin. In the
  upper panel, the solid line is the PDF calculated with the final
  continuum. The solid squares with error bars are the PDF values in
  each flux bin with the associated error estimated by means of the
  jackknife method described in Sect.~\ref{errest}.  The
    short-dashed line, the long-dashed line and the dot-long-dashed
    line represent the PDF calculated by increasing the continuum
    level by 5, 3 and 1.5 per cent, respectively. The dotted line, the
    dot-short-dashed line and the short-dashed-long-dashed line
    represent the PDF calculated by decreasing the continuum level by
    5, 3 and 1.5 per cent, respectively. In the lower panel, and
    following the same legend as above, the various lines display the
    ratio between the PDF calculated by increasing or decreasing the
    continuum and the PDF calculated with the final continuum. The
    errors on the ratios in the lower panel are similar for all curves
    but plotted only once for clarity. }
\label{contpdf}
\end{figure}

\subsection{The effect of the continuum uncertainty}
\label{conteff}

The effects of the systematic uncertainty due to the continuum fitting
procedure are shown in Fig.~\ref{contpdf} for the full 
sample in the redshift bin centered at $z=2.9$. 
In this figure, the
PDF is measured without removing the metal lines and is shown for the
final continuum level (solid curve), as well as the PDF obtained by
increasing and decreasing the continuum level by 5 per cent (dashed
and dotted curves, respectively).  Here the continuum is
conservatively rescaled by 5 per cent since we have seen from
Fig.~\ref{cont} this value is representative of the continuum
uncertainty for most of the spectra in our sample.  In both cases, the
strongest effects on the PDF are visible at high flux values
($F>0.5$), whereas a rescaling of the continuum has a negligible
effect in the lowest flux bins.

The net effect of increasing the continuum by 5 per cent is to remove
pixels at $F \sim 1$, moving the peak of the original PDF to lower
flux bins.  The bins at values $F<0.9$ are also populated with more
pixels than is the case for the final continuum, and the slope of the
PDF between $F\sim 0.1$ and $F\sim 0.9$ tends to increase overall. On
the other hand, by decreasing the continuum by 5 per cent, pixels are
instead moved towards higher flux bins and the PDF peaks at $F=1$.

The determination of the continuum is a very problematic task and
subject to several uncertainties. However, Fig.~\ref{contpdf} shows
how the PDF would vary if we had systematically underestimated or
overestimated the continuum by 5 per cent \emph{for all absorption
  systems}. This seems rather implausible: it is far more likely that
our continuum fitting method underestimates or overestimates the true
continuum in chunks of our spectra, rather than consistently across
the entire data set.\footnote{Note this is not necessarily true 
for moderate to low resolution
and S/N data, which are often normalised assuming a power-
law extrapolation from the relatively unabsorbed spectral regions
uncontaminated by emission lines redwards of the QSO Lyman $\alpha$.
However, more accurate approaches also for this regime,
based on principal component analysis,  have been developed
(e.g. Lee et al. 2012, Paris et al. 2011).}
If this is the case, the
effect of the continuum placement on the PDF will be local (i.e. it
should cause small variations around particular flux values), and will
not affect its global average as strongly as Fig.~\ref{contpdf}
suggests.  Regardless, however, when we turn to obtaining constraints
on the thermal state of the IGM from our PDF measurements later, we
follow Bolton et al. (2008) and Viel, Bolton \& Haehnelt (2009) and
only consider the PDF in the flux range $F=[0.1$--$0.8]$.  This
ignores the part of the PDF which is the most sensitive to continuum
uncertainties, $F>0.8$.

\subsection{Redshift evolution of the PDF}
\label{PDF_z}

\begin{figure*}
\centering
\vspace{0.001cm}
\epsfig{file=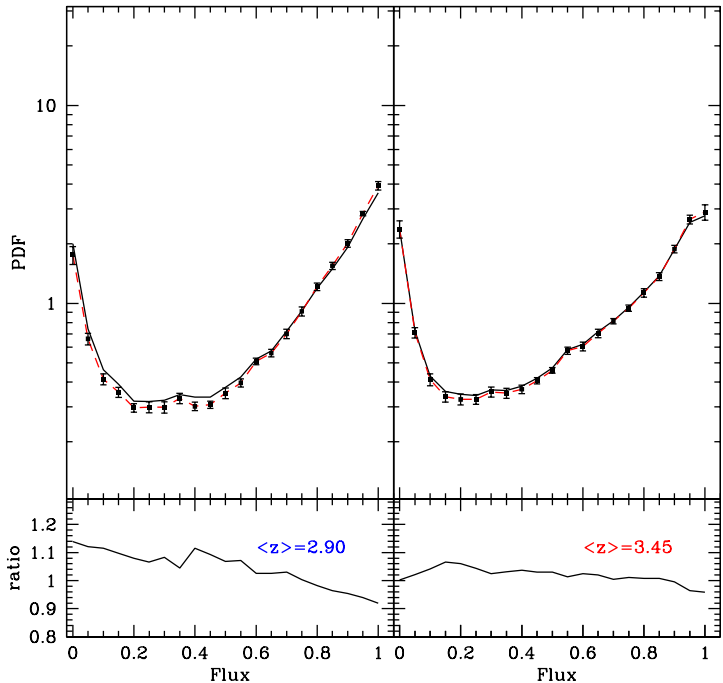,height=10cm,width=10cm}
\caption{The redshift evolution of the PDF with the full sample
  divided into two redshift bins, $\langle z\rangle=2.90$ (left panels)
  and $\langle z\rangle=3.45$ (right panels).  {\it Upper panels}: The
  squares with error bars represent the PDF measured after removing
  the metal line absorption. The dashed lines join the squares and are
  drawn in order to guide the eye.  The solid lines are the PDFs
  computed for the full sample, but including metal absorption lines.
  {\it Lower panels}: The ratios between PDF calculated before and
  after removing the metal lines. }
\label{fig7}
\end{figure*}

Fig.~\ref{fig7} shows the PDF computed for the full sample, divided
into two different redshift bins, centered at redshifts 2.90 and
3.45. This figure also illustrates the effect of metal absorption
lines on the PDF. The ratio between the PDF computed before and after
removing metal absorption lines depends weakly on redshift.

\begin{figure}
\centering
\vspace{0.001cm}
\epsfig{file=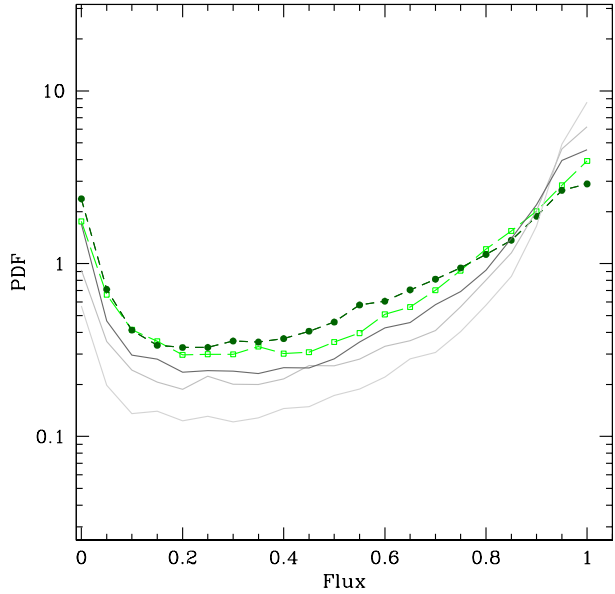,height=8.5cm,width=8.5cm}
\caption{The PDF measured in this work at $\langle z\rangle=2.9$
  (squares and long-dashed line), $\langle z\rangle=3.45$ (circles and
  short-dashed line) compared with the PDF obtained by Kim et
  al. (2007) at $\langle z\rangle=2.07$ (light grey), $\langle
  z\rangle=2.52$ (grey) and $\langle z\rangle=2.94$ (dark grey). }
\label{fig9}
\end{figure}

The PDF for the full sample, divided into two redshift bins, is
provided in Table~\ref{tab2}. We tabulate the PDF computed for (1) all
pixels, (2) after removing contamination from metal absorption lines
and (3) after removing both metal absorption lines and Lyman limit
systems; the latter has been used for the comparison with simulation
results as described in Sect.~\ref{simul}. In Fig. \ref{fig9}, the
PDFs excluding metals are compared to the PDFs obtained by K07 at
lower redshift.  For purposes of clarity, and to better focus on the
evolution of the PDF, here the error bars are omitted.  As already
noted, the spectra at higher redshift are characterised by lower
average flux levels.  In terms of the PDF, this translates into a
greater number of pixels with a low fluxes, resulting in a shallower
distribution toward higher redshift.  This trend is clearly visible in
Fig.\ref{fig9}.

It is also worth noting that a different method to remove metal
absorption lines was used by K07.  In the work of K07, pixels
belonging to spectral regions including metals were not excised. The
PDF was instead calculated by using the spectra fitted with VPFIT,
including only Lyman $\alpha$ absorption lines.  The use of these
different methods could cause small differences between the
PDFs. However, we have checked that these differences are within the
error bars estimated with the jackknife method, and the results shown
in Fig. \ref{fig9} are unlikely to be significantly affected by the
different methods used to remove metals from the QSO spectra.

\begin{figure}
\centering
\vspace{0.001cm}
\epsfig{file=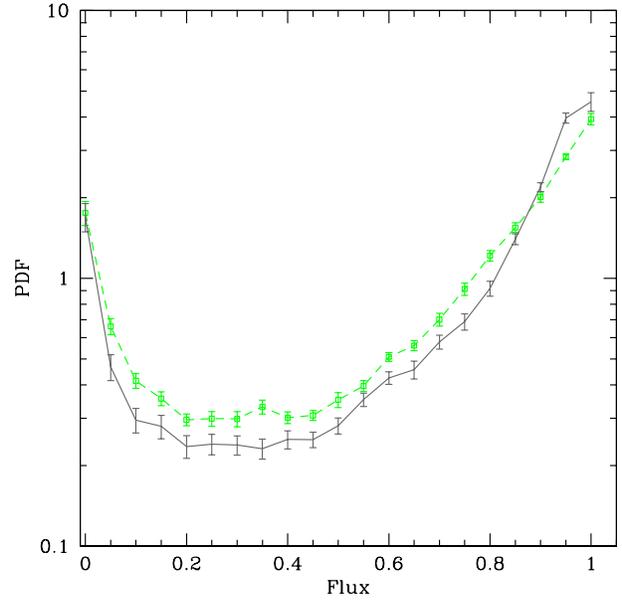,height=8.5cm,width=8.5cm}

\caption{The PDF measured in this work in the lowest redshift bin
  (squares and long-dashed curve) plotted with error bars compared to
  the PDF obtained by Kim et al. (2007) at $\langle z\rangle=2.94$
  (dark grey curve). }
\label{fig9a}
\end{figure}

\begin{figure}
\centering
\vspace{0.001cm}
\epsfig{file=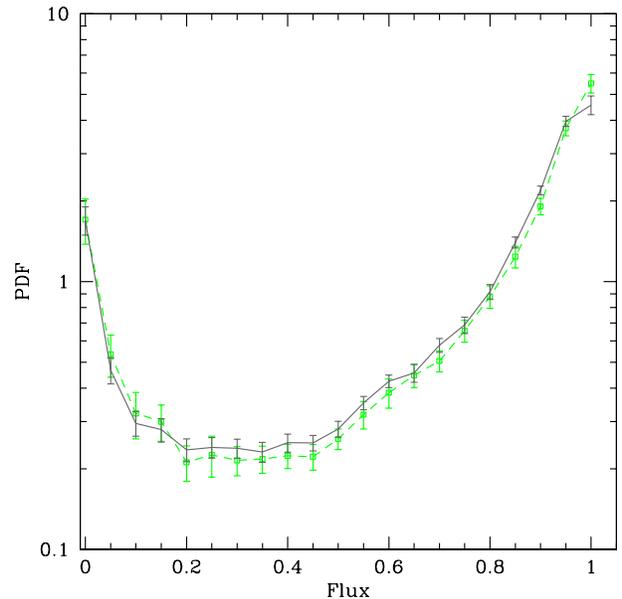,height=8.5cm,width=8.5cm}
\caption{The flux PDF measured by K07 at $\langle z\rangle=2.94$ (dark
  grey curve) plotted with error bars compared to the PDF measured
  from the two spectra in our sample, Q0055-269 and PKS2126-158, used
  in the K07 measurement (open squares and long dashed curve).  This
  comparison uses pixels in the same wavelength range adopted K07.}
\label{fig9b}
\end{figure}

In addition, in Fig. \ref{fig9} there is a difference between our PDF
measured in the lowest redshift bin and the PDF obtained by K07 in
their highest redshift bin, i.e. at $\langle z\rangle=2.94$. 
This difference is better illustrated in Fig~\ref{fig9a}, where we
report our PDF calculated at $\langle z\rangle=2.90$ and the PDF
measured by K07 at $\langle z\rangle=2.94$.  Our PDF is higher than
the K07 measurement by a factor ranging from 1.1 to 1.5 in the flux
range $0.05 \le F \le 0.8$.  Note that two of the
QSO spectra used in this study were also used in the PDF measurements
presented by K07 (Q0055-269 and PKS2126-158). These objects contribute
significantly to the highest bin of K07, as well as to our lowest
redshift bin.  
As a sanity check, we have re-calculated the PDF with the two spectra
of Q0055-269 and PKS2126-158 only, by using the pixels in the same
wavelength range adopted by K07. The results of this test are visible
in Fig.~\ref{fig9b}, along with jackknife error bars: 
once the same wavelength
ranges as those of K07 for Q0055-269 and PKS2126-158 are taken into
account, the two PDFs are consistent with each other within the
uncertainties. This result tells us that the discrepancy between our
PDF and K07's is not due to a mistake in the computation of the PDF.

Since we expect that the flux PDF in a given redshift range does not
depend on the sample of QSO spectra used to compute it, the observed
discrepancy could be due to the modest size of the sample. If this is
the case, our result should be more representative of the true value
of the PDF since the sum of the $\Delta z$ of our QSO spectra
contributing  to the considered bin is larger than $\sim$ 
50 per cent with
respect to K07. We note also that the average redshift of the pixels
belonging to our lowest redshift bin is larger than the corresponding
value for K07’s highest redshift bin ($\langle z_{pixel}\rangle= 3.03$ 
versus 2.91,
respectively). 
This difference could contribute to the fact that, as
seen in Fig.\ref{fig9}, our PDF at $\langle z\rangle=2.90$ is 
higher than K07’s PDF at at $\langle z\rangle=2.94$. 

Summarizing, while the evolution with redshift 
of the flux PDF is evident from Fig.\ref{fig9}, more observations of QSOs at
$z\sim 3.3$ could be needed to establish the true level of the flux PDF at
$z\sim 2.9$.

\begin{figure}
\centering
\vspace{0.001cm}
\epsfig{file=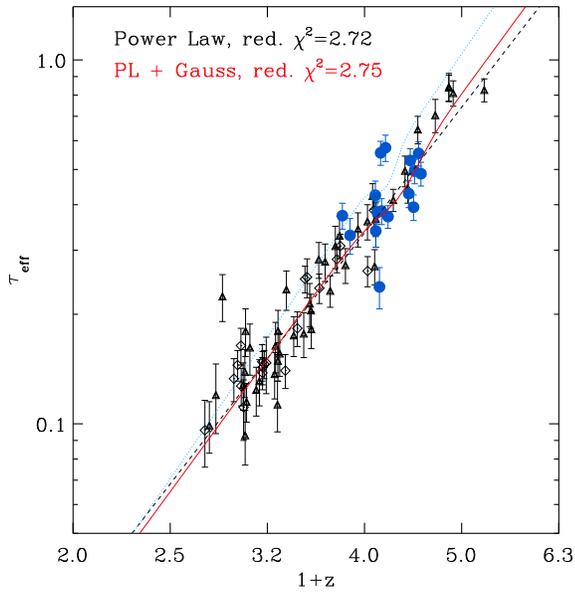,height=8.5cm,width=8.5cm}
\caption{The redshift evolution of the Lyman $\alpha$ forest effective
  optical depth. The solid circles are the measurements from the
  present work.  The triangles are the data from Schaye et al. (2003)
  and the diamonds are from Kim et al. (2007).  The dashed line is the
  best fit power-law and the solid line the power-law plus Gaussian
  which best fits the whole sample of data (see text for details).
  The dotted line is the best fit power-law plus Gaussian obtained by
  Faucher-Giguere et al. (2008) from the study of a sample of
  high-resolution spectra.}
\label{fig10}
\end{figure}

\begin{figure}
\centering
\vspace{0.001cm}
\epsfig{file=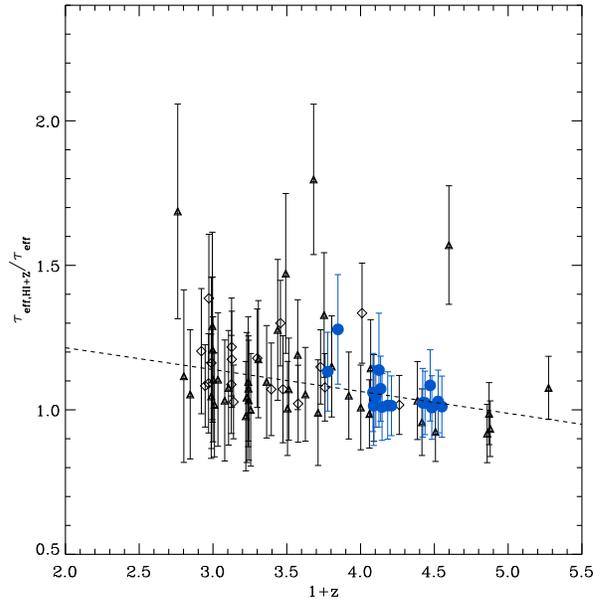,height=8.5cm,width=8.5cm}
\caption{Redshift evolution of the ratio between the optical depth
  computed without removing the metal absorption and after the removal
  of the metal lines.  The solid circles are the values computed in
  the present work.  The triangles are the data by Schaye et
  al. (2003) and the diamonds are the data by Kim et al. (2007).  The
  dashed line represents a straight-line fit to the data, $y=a \, x
  +b$, with $a=1.37$ and $b=-0.076$.}
\label{fig11}
\end{figure}

\subsection{Redshift evolution of the HI optical depth}
\label{tau}

A quantity of interest in cosmology is the effective optical depth of
the Lyman $\alpha$ forest, which is also sensitive to the thermal and
ionization state of the IGM at high redshift.  If $\langle
F\rangle_{\rm HI} = \exp\left(-\tau_{\rm eff}\right)$ represents the
average normalized flux measured from the Lyman $\alpha$ forest, the
effective optical depth may be defined as $\tau_{\rm eff}\equiv -\ln
\langle F\rangle_{\rm HI}$.  The effective optical depth is therefore
related to the average flux in all pixels in the \lya forest, and
provides different and complementary measure to the PDF, which is a
differential measurement.

In order to better probe the redshift range investigated in this paper
and to increase the amount of information which can be extracted from
our sample, we have divided each of the eight spectra analysed in this
work into two redshift bins.  In each bin we have then calculated the
effective optical depth. In Tab.~\ref{tab3}, we show the effective
optical depth for our sample, computed both before and after removing
metal contamination.  We have calculated $\tau_{\rm eff}$ by
considering the flux in the wavelength ranges reported in
Tab.~\ref{tab3}, after removing a fraction of pixels as described in
section \ref{voigt} in order to avoid the proximity effect.  Note also
that since we wish to measure the HI effective optical depth, removing
metal lines is again very important for obtaining a precise estimate
of $\tau_{\rm eff}$ which can be compared to other sets of data or with
simulations.

The errors on $\tau_{\rm eff}$ are computed by means of a bootstrap
method, as described in McDonald et al. (2000), Schaye et al. (2003)
and K07.  We have divided each spectrum into $N$ chunks of 100 pixels,
corresponding to $5$ \AA.  A total of 500 bootstrap realizations were
then performed by randomly selecting the $N$ chunks with
replacement. The standard deviation from the mean $\tau_{\rm eff}$ is
the error reported in Tab.~\ref{tab3}.  In the last column of
Tab.~\ref{tab3} we also give the percentage contribution of metal
absorption lines to $\tau_{\rm eff}$, which can be calculated as:
\begin{equation}
{\mathrm metals~(per~cent)} = (\frac{\tau_{\rm eff,HI+Z}}{\tau_{\rm eff}} -1) \cdot 100.
\end{equation}
The metal contamination varies from 1 to 28 per cent; we will discuss
its redshift dependence in more detail later on in this section.

In Fig.~\ref{fig10}, we show the evolution of Lyman $\alpha$ forest
optical depth, $\tau_{\rm eff}$, measured in this work after the
removal of metal lines, compared to previous estimates by various
authors.  The dotted line is the best fit obtained by Faucher-Giguere
et al. (2008) from a sample of 86 high-resolution, high-S/N ESI and
HIRES quasar spectra. These authors also find evidence for a deviation
of $\tau_{\rm eff}$ from a power-law at $z\sim 3.2$, consistent with
the previous result of Bernardi et al. (2003) obtained using SDSS data
and a different methodology (see also P{\^a}ris et al. 2011). 
The solid line is our best fit to all of
the data (i.e. those of Schaye et al. 2003, Kim et al. 2007 and the
present work) obtained by means of the relation
\begin{equation}
\tau_{\rm eff}=A(1+z)^{B}+C \exp \{-\frac{[(1+z)-D]^2}{2E^2}\},
\end{equation}
which is a modified power-law including a Gaussian component to take
into account the presence of a deviation, as in Faucher-Giguere et
al. (2008). In comparison, the dashed line represents the best fit to
the whole data ensemble for a power-law only. The reduced $\chi^2$ are
reported in Fig.~\ref{fig10}; we find that a power law provides a
better fit to the data. In the case of a power law plus Gaussian, the
best fit was obtained with $A=0.0023, B=3.64, C=-0.05, D=4.36,
E=0.21$, i.e. with parameters associated with a very tiny deviation
from a power law.  Note also that the reduced $\chi^2$ reported
  in Fig ~\ref{fig10} are in general larger than the values reported
  in table 5 of Faucher-Giguere et al. (2008).  However, the data used
  by Faucher-Giguere et al. (2008) are more homogeneous and show a
  lower dispersion, hence it seems reasonable for their fits to show
  lower $\chi^2$ values.  From these results, we conclude that, from
our study of the redshift evolution of the effective optical depth, we
do not find any strong evidence for a deviation from a power law.  The
best fit is expressed by the power law $\tau_{\rm eff}=0.0028 \pm 0.0003\, 
  (1+z)^{3.45 \pm 0.08}$, 
in very good agreement with the previous best fit
provided by K07: $\tau_{\rm eff}=0.0023\pm 0.0007 (1+z)^{3.65\pm
  0.21}$.

In Fig.~\ref{fig11}, we show the redshift evolution of the ratio
between the effective optical depth measured before and after removing
metal absorption lines.  This plot illustrates the effect of metal
absorption on the Lyman $\alpha$ forest effective optical depth as a
function of redshift.  Our data are plotted together with the
measurements from K07 and Schaye et al. (2003).  The dashed line
represents a straight-line fit to the data, $y=a \, x +b$. Our best
fit values are $a=1.36$ and $b=-0.07$.  The weak correlation shown in
Fig.~\ref{fig11} again highlights the previously discussed redshift
dependence of the metal line contamination in the \lya forest, which
is consistent with a greater amount of metal absorption toward lower
redshift.

\section{The simulations}
\label{simul}
In this section, we now turn to describe the suite of hydrodynamic
simulations used to obtain constraints on the thermal state of the IGM
at redshift $z\,{}^>_{\rm \sim}\,3$ from our PDF measurements.  
  Although the basic setup of the simulations is similar to earlier
  work by Bolton et al. (2008) and Viel et al. (2009), 
  we now use new, 
  higher resolution runs to explore a wider thermal history parameter
  space compared with those used by Viel et al. (2009).  The
simulations were performed using the parallel Tree-PM smoothed
particle hydrodynamics (SPH) {\small GADGET-2} code (Springel 2005)
and its recently updated version {\small GADGET-3}. All the
simulations were started at $z=99$, with initial conditions generated
using the transfer function of Eisenstein \& Hu (1999). The
gravitational softening length was chosen to be equal to 1/30th of the
mean linear interparticle spacing and star formation was included
using a simplified prescription which converts all gas particles with
temperature $T<10^5$ K and overdensity $\Delta = \rho / \langle \rho
\rangle > 10^3$ into collisionless stars. 
\begin{table*}
\label{tab4}
\caption{Summary of the hydrodynamical simulations used in this
  work. Column 1, simulation name: note the distinction between subset
  A (high resolution set) and the subset B (low resolution set);
  column 2, number of gas particles; column 3, mass of each gas
  particle; column 4, matter density parameter.  In all models we
  assume a flat universe with $\Omega_{\rm \Lambda}=1-\Omega_{\rm
    0m}$. However, $\Omega_{\rm b}h^2=0.023$ for the subset $A$ and
  $\Omega_{\rm b}h^2=0.024$ for the subset $B$ of our simulations;
  column 5, primordial power spectrum index; column 6, Hubble
  constant; column 7, fluctuation amplitude at a scale of 8 $h^{-1}$
  Mpc; column 8, slope of the temperature-density ($T-\Delta$)
  relation, where $T(\Delta)=T_{\rm 0}\,\Delta^{\gamma-1}$; column 9:
  gas temperature at the mean density, in units of $10^3$ K, at
  $z=3.25$; column 10: further details on the simulations.}
\begin{tabular}{llccccccccl}
\hline 

& Run & $N_{\rm GAS}$ & $m_{\rm GAS}$ & $\Omega_{\rm 0m}$ & $n_{\rm
  s}$ & $H_{\rm 0}$ & $\sigma_{\rm 8}$ & $\gamma$ & $T_{\rm 0}$
($10^3$ K) & Notes \vspace*{0.1cm}\\

& & & {\small ($h^{-1}$) M$_{\odot}$} & & & {\small (km/s/Mpc)} & & &
{\small ($z=3.25$)} & \\
 
\hline 

& A1$_{\rm REF}$ & $512^3$ & $9.2 \times 10^4$ & 0.26 & 0.96 & 72 &
0.80 & 1.0 & 19.42 & Fiducial high res. model \\

& A1$_{\rm CLD}$ & $512^3$ & $9.2 \times 10^4$ & 0.26 & 0.96 & 72 &
0.80 & 1.0 & 14.93 & Cold run \\

& A1$_{\rm HOT}$ & $512^3$ & $9.2 \times 10^4$ & 0.26 & 0.96 & 72 &
0.80 & 1.0 & 23.88 & Hot run \\

& A1$_{\rm LW\gamma}$ & $512^3$ & $9.2 \times 10^4$ & 0.26 & 0.96 &
72 & 0.80 & 0.7 & 19.96 & Low $\gamma$ \\

& A1$_{\rm HI\gamma}$ & $512^3$ & $9.2 \times 10^4$ & 0.26 & 0.96 &
72 & 0.80 & 1.3 & 19.07 & High $\gamma$ \\

& B1$_{\rm REF}$ & $256^3$ & $7.4 \times 10^5$ & 0.26 & 0.95 & 72 &
0.85 & 1.0 & 23.79 & Low res. reference model\\

& B1$_{\rm LS8}$ & $256^3$ & $7.4 \times 10^5$ & 0.26 & 0.95 & 72 &
0.80 & 1.0 & 23.64 & Low $\sigma_{\rm 8}$ \\

& B1$_{\rm HS8}$ & $256^3$ & $7.4 \times 10^5$ & 0.26 & 0.95 & 72 &
0.90 & 1.0 & 23.95 & High $\sigma_{\rm 8}$ \\

& B1$_{\rm LNs}$ & $256^3$ & $7.4 \times 10^5$ & 0.26 & 0.90 & 72 &
0.85 & 1.0 & 23.67 & Low $n_{\rm s}$ \\

& B1$_{\rm HNs}$ & $256^3$ & $7.4 \times 10^5$ & 0.26 & 1.00 & 72 &
0.85 & 1.0 & 23.93 & High $n_{\rm s}$ \\

& B1$_{\rm LOm}$ & $256^3$ & $7.4 \times 10^5$ & 0.22 & 0.95 & 72 &
0.85 & 1.0 & 24.55 & Low $\Omega_{\rm 0m}$ \\

& B1$_{\rm HOm}$ & $256^3$ & $7.4 \times 10^5$ & 0.30 & 0.95 & 72 &
0.85 & 1.0 & 23.12 & High $\Omega_{\rm 0m}$ \\

& B1$_{\rm LH0}$ & $256^3$ & $7.4 \times 10^5$ & 0.26 & 0.95 & 64 &
0.85 & 1.0 & 25.40 & Low $H_{\rm 0}$ \\

& B1$_{\rm HH0}$ & $256^3$ & $7.4 \times 10^5$ & 0.26 & 0.95 & 80 &
0.85 & 1.0 & 22.35 & High $H_{\rm 0}$ \\

\hline
\end{tabular}
\end{table*}

In Table~\ref{tab4} we summarize the parameters used in the
cosmological simulations. We use two different sets of
simulations. Set A includes high resolution runs and, among them, the
fiducial run for this paper: A1$_{\rm REF}$. Each simulation has a box
size of 10 $h^{-1}$ Mpc comoving and contains 2 $\times$ 512$^3$ dark
matter and gas particles. The model chosen is a flat $\Lambda$CDM
cosmology with parameters: $\Omega_{\rm 0m}=0.26$, $\Omega_{\rm
  \Lambda}=0.74$, $\Omega_{\rm b}h^2=0.023$, $n_{\rm s}=0.96$, $H_{\rm
  0}=72$ km s$^{-1}$ Mpc$^{-1}$ and $\sigma_{\rm 8}=0.80$, which are
in agreement with recent studies of the cosmic microwave background
(Komatsu et al. 2009; Reichardt et al. 2009; Jarosik et al. 2011).

We use set A to explore variations in the thermal state of the gas. A
spatially uniform ultra-violet background (UVB) produced by quasars
and galaxies (Haardt \& Madau 2001) was adopted, where hydrogen is
reionized at $z=9$. 

The gas in the simulations is assumed to be optically thin and in
ionization equilibrium with the UVB (i.e. photoheating by the UVB is
balanced by cooling due to adiabatic expansion). In these physical
conditions, gas with overdensity $\Delta\,{}^<_{\rm \sim}\,10$ 
follows a tight power-law 
temperature-density relation: $T=T_{\rm
  0}\,\Delta^{\gamma-1}$ ( Hui \& Gnedin
1997; Valageas, Schaeffer \& Silk 2002), 
where $T_{\rm 0}$ is the temperature of
the IGM at the mean density.
In order to obtain various thermal histories,
the Haardt \& Madau (2001) photo-heating rates $\epsilon_{\rm
  i}^{\small \rm HM01}$ were rescaled by different values:
$\epsilon_{\rm i}=k\,\Delta^q\epsilon_{\rm i}^{\small \rm HM01}$,
where $i=$[HI, HeI, HeII]. Our fiducial simulation A1$_{\rm REF}$ has
parameters $k=2.20$ and $q=-1.00$ (which produces an isothermal
$T-\Delta$ relation). Runs A1$_{\rm CLD}$ and A1$_{\rm HOT}$ also have
$q=-1.00$ (still an isothermal $T-\Delta$ relation), but $k=1.45$
(producing lower $T_{\rm 0}$) and $k=3.10$ (higher $T_{\rm 0}$),
respectively. Runs A1$_{\rm LW\gamma}$ and A1$_{\rm HI\gamma}$ both
have $k=2.20$, but $q=-1.60$ (resulting in an inverted $T-\Delta$
relation with $\gamma=0.7$) and $q=-0.45$ (flattened $T-\Delta$
relation with $\gamma=1.3$), respectively.

Simulation set B includes lower resolution runs that are intended to
explore the dependency of the PDF on cosmological parameters. These
simulations have a box size of 10 $h^{-1}$ Mpc comoving and contain 2
$\times$ 256$^3$ dark matter and gas particles. Again, the reference
cosmological model (see run B1$_{\rm REF}$ in Tab.~\ref{tab4}) is a
flat $\Lambda$CDM model, but now the parameters are slightly different: 
$\Omega_{\rm 0m}=0.26$,
$\Omega_{\rm \Lambda}=0.74$, $\Omega_{\rm b}h^2=0.024$, $n_{\rm
  s}=0.95$, $H_{\rm 0}=72$ km s$^{-1}$ Mpc$^{-1}$ and $\sigma_{\rm
  8}=0.85$.  As shown in Tab.~\ref{tab4}, the cosmological parameters
varied are $\sigma_{\rm 8}$ (runs B1$_{\rm LS8}$ and B1$_{\rm HS8}$),
$n_{\rm s}$ (runs B1$_{\rm LNs}$ and B1$_{\rm HNs}$), $\Omega_{\rm
  0m}$ (runs B1$_{\rm LOm}$ and B1$_{\rm HOm}$) and $H_{\rm 0}$ (runs
B1$_{\rm LH0}$ and B1$_{\rm HH0}$).  Note that this second set of
simulations has a slightly different $\sigma_8$ value than the first
set, however, we do take this into account in the analysis described
below. We refer the reader to Becker et al. (2011), Bolton et
al. (2008) and Viel, Bolton \& Haehnelt (2009) for further details on
the simulations.

\section{ANALYSIS OF THE SIMULATIONS}

\subsection{Synthetic spectra and PDFs}

Given the positions, velocities, densities and temperatures of all the
SPH particles at a given redshift, spectra along lines-of-sight (LOSs)
through the simulation boxes were computed following the procedure
described by Theuns et al. (1998). The interested reader can find more
details about this procedure in Section 5 of Tescari et al. (2011).
Two different sets of synthetic spectra were constructed: the first
set includes two redshift bins at $z=2.95$ and $3.48$ and the second
set includes three redshift bins at $z=2.90, 3.25$ and $3.55$.  After
extracting the spectra along random LOS through the cosmological box
at redshift $z$, we rescaled all the HI optical depths by a constant
factor, so that their mean value was equal to the HI effective optical
depth, $\tau_{\rm eff}$, given by the Kim et al. (2007) fit:
\begin{equation}
\label{eq_6}
\tau_{\rm eff} =(0.0023\pm0.0007)(1+z)^{3.65\pm 0.21}.
\end{equation}
This rescaling ensures that our spectra match the observed mean
normalized flux of the Lyman $\alpha$ forest at the appropriate
redshift: $\langle F\rangle_{\rm HI,obs} = \exp\left(-\tau_{\rm
    eff}\right)$. The spectra are then convolved with a Gaussian of 7
km s$^{-1}$ FWHM and rebinned on to pixels of width 0.05 \AA. Finally,
in order to have realistic spectra to compare with observations, we
add Gaussian distributed noise with total signal to noise/pixel
(i.e. standard deviation of the Gaussian at $F=1$) equal to 60 and
read out S/N equal to 100.

\begin{figure*}
\centering
\epsfig{file=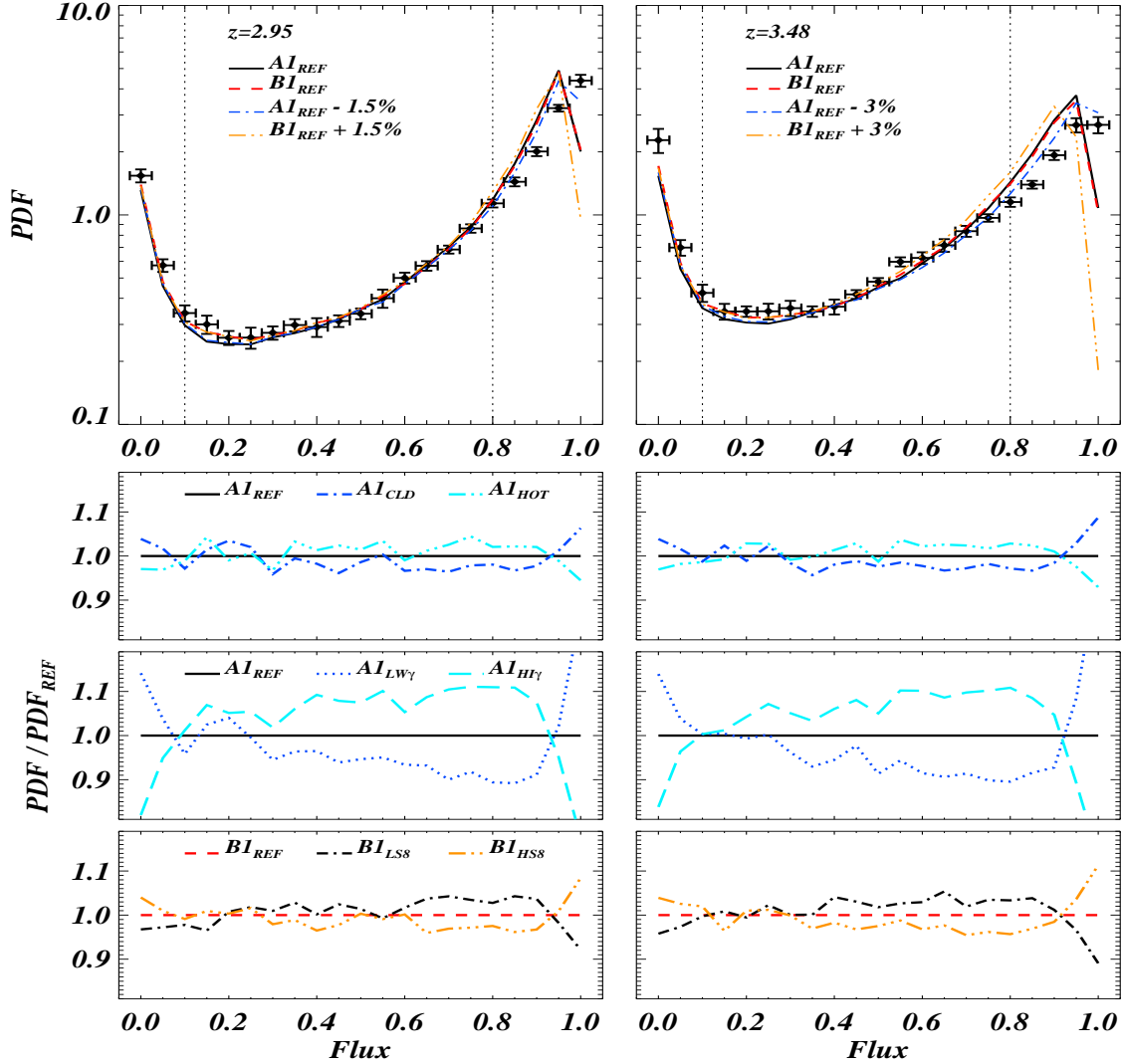,height=16cm,width=15cm}
\caption{Simulated PDFs at redshift $z=2.95$ (left-hand side) and
  $z=3.48$ (right-hand side). {\it Upper two panels}: Our two
  reference simulations are displayed in both plots: run A1$_{\rm
    REF}$ (black solid curves) and run B1$_{\rm REF}$ (red dashed
  curves). The black data points show the PDF measured from the full
  observational sample after the removal of metals and LLS (Tab. 2,
  last two columns). The blue dot-dashed curves and orange triple
  dot-dashed curves are the PDFs obtained with $\pm1.5\%$ (left-hand
  side) and $\pm3\%$ (right-hand side) continuum errors added to the
  spectra from the runs A1$_{\rm REF}$ and B1$_{\rm REF}$,
  respectively. The two vertical dotted lines mark the region inside
  which we perform the fit to the observational data in this
  work. {\it Lower six panels}: the ratio between PDFs obtained from
  simulations with different values of astrophysical and cosmological
  parameters, and the corresponding reference PDF$_{\rm
    REF}$. Variations in the following parameters are explored:
  $T_{\rm 0}$ (first row), $\gamma$ (second row) and $\sigma_{\rm 8}$
  (third row).}
\label{fig_pdf_sim}
\end{figure*}

  The dependence of the simulated flux PDF on $T_{0}$, $\gamma$ and
  $\sigma_{8}$ is illustrated in Fig.~\ref{fig_pdf_sim}.  The two
  upper panels show the PDFs extracted from our two fiducial
  simulations: the high resolution run A1$_{\rm REF}$ (black solid
  curve) and the low resolution run B1$_{\rm REF}$ (red dashed curve)
  at redshift $z=2.95$ (left-hand side) and $z=3.48$ (right-hand
  side). Each model is compared to the PDF of the full observational
  sample measured after the removal of metals and LLS (data points
  with error bars). The simulated PDFs in this instance have been
  obtained by rescaling the value of the effective optical depth,
  $\tau_{\rm eff}$, in order to minimize the $\chi^2$ statistic.  In
  the analysis presented in Sec. 5.2, we will only compare our
  simulations to flux values in the range $[0.1-0.8]$, which is less
  sensitive to possible continuum errors.  To demonstrate this, the
  simulated PDFs from run A1$_{\rm REF}$ (blue dot-dashed lines) and
  B1$_{\rm REF}$ (orange triple dot-dashed lines) with additional
  $\pm1.5\%$ (left-hand side) and $\pm3\%$ (right-hand side) continuum
  errors are overplotted. Even in the worst case of $\pm3\%$ errors,
  the simulated PDFs are consistent in the range $F=[0.1-0.8]$. 

  The lower six panels of Fig.~\ref{fig_pdf_sim} show the ratio
  between PDFs obtained from simulations with different values of
  astrophysical and cosmological parameters, and the corresponding
  reference PDF$_{\rm REF}$ (calculated by using either A1$_{\rm REF}$
  or B1$_{\rm REF}$), at the same two redshifts. We vary the following
  parameters: $T_{\rm 0}$ (first row), $\gamma$ (second row) and
  $\sigma_{\rm 8}$ (third row). It is clear from the figure that
  variations of $\gamma$ have the largest effect on the simulated PDF
  (see also Fig. 2 of Bolton et al. 2008 and associated discussion).

\begin{figure*}
\centering
\vspace{0.001cm}
\epsfig{file=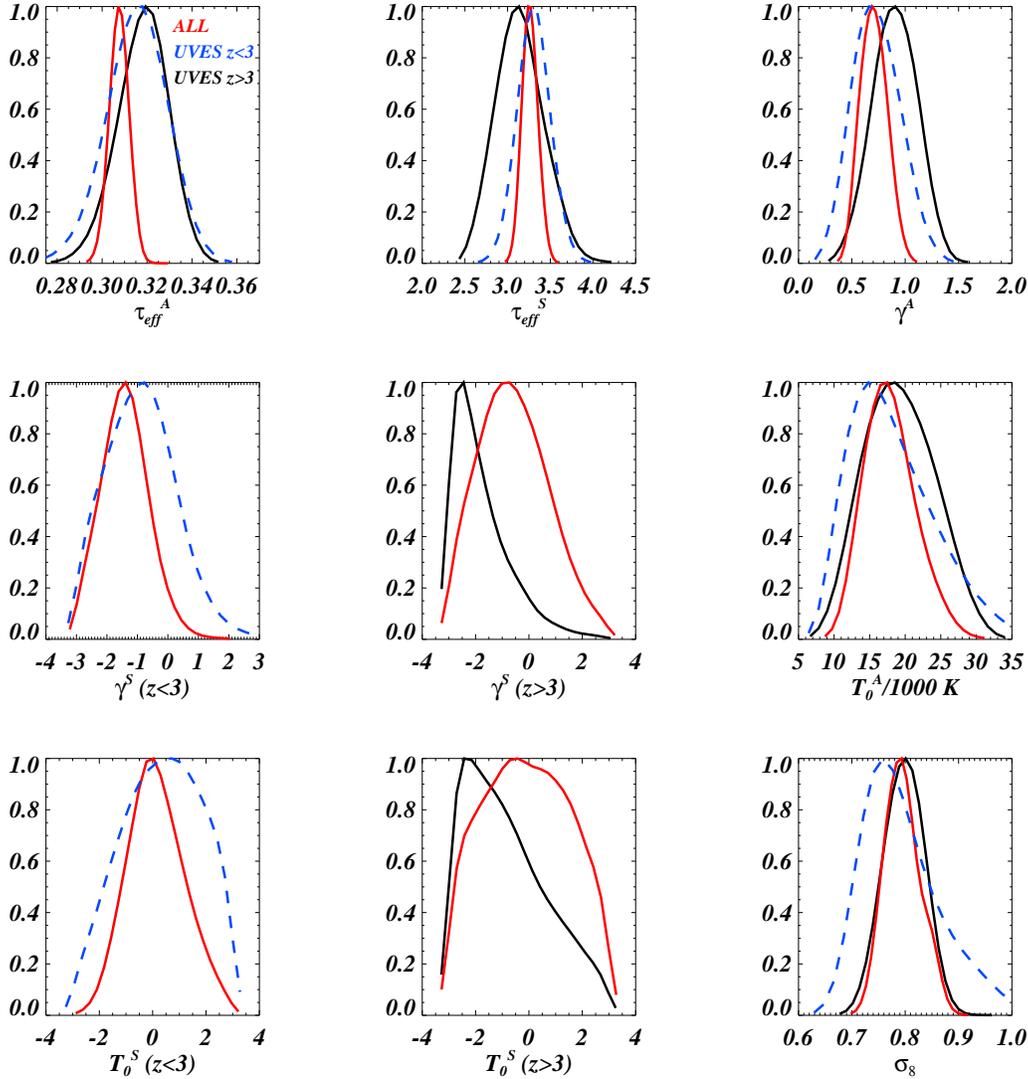,height=15cm,width=15cm}
\caption{One-dimensional marginalized constraints on the cosmological
  and astrophysical quantities used in the analysis (see text for
  details of the parametrization). The blue curves indicate the
  constraints obtained by using the PDF measurements at $z<3$
  presented by K07 (see also the analysis of Viel, Bolton \& Haehnelt
  2009), black-continuous curves refer to the present work, while the
  joint constraints for both this data set and the measurements of K07
  are shown as continuous red curves.}
\label{fig12}
\end{figure*}

\begin{table*}
\label{tab5}
\caption{$1-\sigma$ parameter constraints from the sample of QSOs
  analysed in $a)$ this work and $b)$ the joint sample composed of
  this work and the K07 PDF data.}
\begin{tabular}{llcccccccccc}

\hline 

& & $\tau_{\rm eff}^{\rm A}$ & $\tau_{\rm eff}^{\rm S}$ & $\gamma^{\rm
  A}$ & $\gamma^{\rm S} (z<3) $ & $\gamma^{\rm S} (z>3)$ &$T_{\rm
  0}^{\rm A}$ [K] &$T_{\rm 0}^{\rm S}$ ($z<3$) & $T_{\rm
  0}^{\rm S}$ ($z>3$) &  $\sigma_{\rm 8}$ & \\

\hline 

& this work & 0.322 & 3.15 & 0.90 & -- & -1.8 & 19250 & -- & -0.8 & 0.8
& \\

& &$\pm$0.012 & $\pm$0.28 & $\pm$0.21 &  & $\pm$1.0 & $\pm$4800 & &
$\pm$1.5 & $\pm$0.04 & \\

& & & & & & & & & & & \\

& this work + K07 data & 0.308 & 3.26 & 0.70 & -1.4 & -0.6 & 17900 &
0.1 & -0.1 & -- & \\

&  & $\pm$0.005 & $\pm$0.10 & $\pm$0.12 & $\pm$0.7 & $\pm$1.3 &
$\pm$3500 & $\pm$1.0 & $\pm$1.6 & & \\

\hline
\end{tabular}
\end{table*}

\subsection{Constraints on cosmological and astrophysical parameters}

We next use the synthetic spectra to construct flux PDFs which may be
compared to our observational measurements.  We explore the effect of
varying the following cosmological and astrophysical parameters on the
PDF: $\sigma_{\rm 8}$, $n_{\rm s}$, $\Omega_{\rm 0m}$ and $H_{\rm 0}$
for the cosmological parameters, and $T_{\rm 0}$ and $\gamma$ for the
IGM thermal history.  Note, however, that most of the cosmological
parameters are only weakly constrained by the flux PDF.  The largest
effect is induced by using a different value of $\sigma_8$
(e.g. Bolton et al. 2008).  We will therefore present marginalized
cosmological constraints for the amplitude of the matter power
spectrum only.  The astrophysical parameters $T_{\rm 0}$ and $\gamma$
are parameterized as a broken power-law around $z=3$, in order to
investigate possible changes in the thermal state around HeII
reionization: $T_{0}=T_{0}^{\rm A}(z=3)\times[(1+z)/4]^{T_{0}^{\rm
    S}(z)}$ and $\gamma=\gamma^{\rm
  A}(z=3)\times[(1+z)/4]^{\gamma^S(z)}$ (Viel et al. 2009).  The
effective optical depth evolution is parameterized $\tau_{\rm
  eff}=\tau_{\rm eff}^{\rm A}(z=3)\times[(1+z)/4]^{\tau_{\rm eff}^{\rm
    S}(z)}$.

In order to explore our parameter space efficiently with our limited
number of simulations, we Taylor expand the flux PDF and compute
derivatives to second order using two simulations (plus the reference
run) for each parameter.  This closely follows the procedure described
in Viel \& Haehnelt (2006), where this Taylor expansion method was
used in order to explore constraints derived from the Sloan Digital
Sky Survey flux power spectrum (McDonald et al. 2006).  We start with
the A1$_{\rm REF}$ simulation and proceed to calculate the $\chi^2$ of
our models by varying all the parameters that alter the flux PDF,
enabling us to expand around the fiducial starting model.  The
expansion method allows us to explore the parameter space close to the
reference model with an accurate set of hydrodynamic simulations, even
if the full parameter space cannot be probed in this way with the same
high accuracy.  We impose weak priors on the effective optical depth
evolution, but both its amplitude and slope are constrained strongly
by the data, so the priors do not affect our results. In order to
present a conservative analysis, we also decide to consider only the
values of the flux in the range $F=[0.1-0.8]$, as this range should
not be affected much by the errors made in the continuum fitting
procedure (e.g. K07; Bolton et al. 2008, and see
Fig.~\ref{contpdf}). Furthermore, when estimating the covariance
between the PDF data points we use the covariance extracted from the
simulations for the non-diagonal terms, which is less noisy (e.g. Lidz
et al. 2006).

\begin{figure}
\vspace{0.2cm}
\centering
\epsfig{file=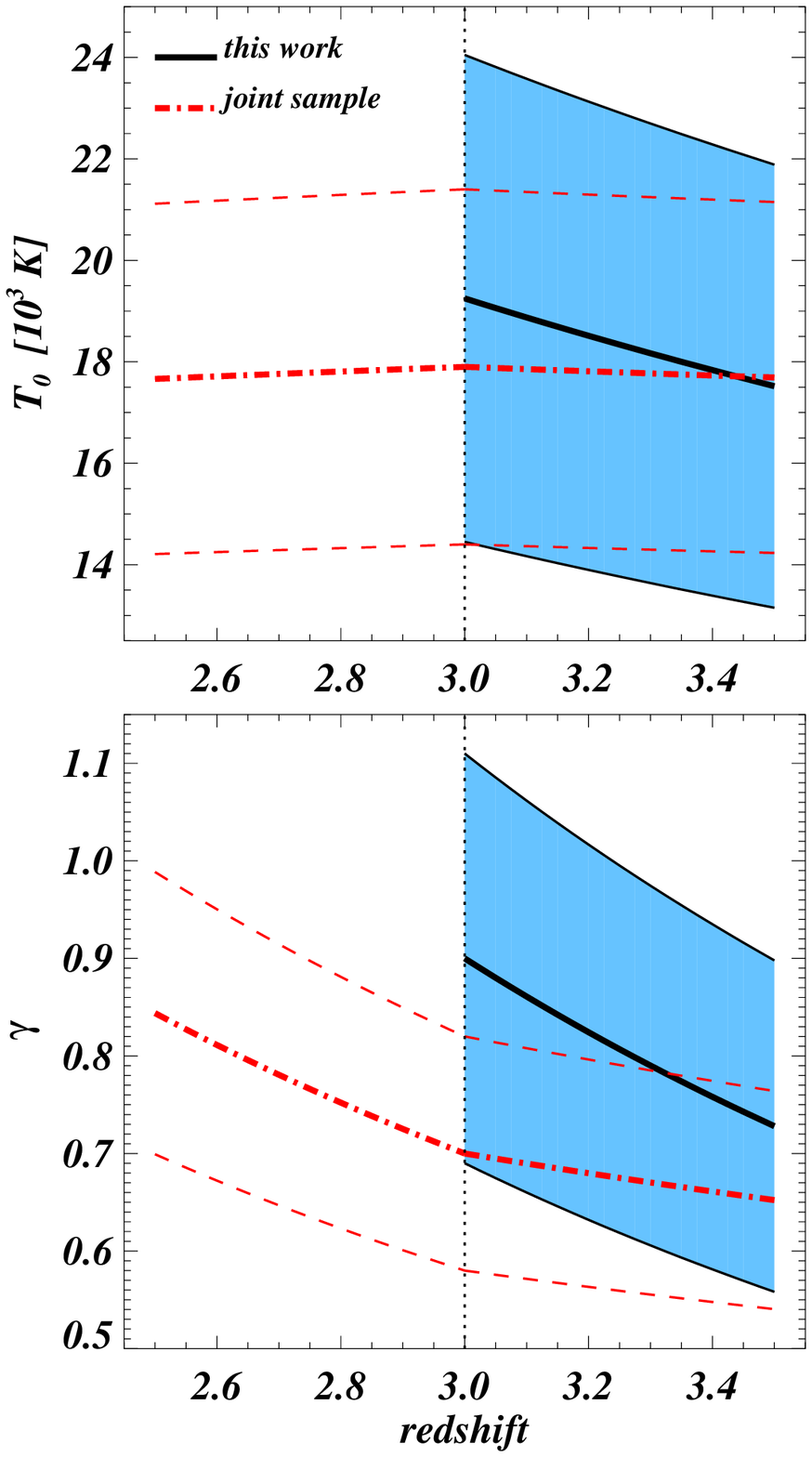,height=11cm,width=7cm}
\caption{Redshift evolution of the temperature at the mean density,
  $T_{0}$, ({\it upper panel}), and the slope of the
  temperature-density relation, $\gamma$, ({\it lower panel)} inferred
  from our analysis of the flux PDF. The black solid lines are the
  results obtained from this work ($z>3$), with the shaded blue
  regions showing the associated errors (see text for details). The
  red dot-dashed and lines are the results obtained considering the
  joint sample composed of this work and the K07 data. The associated
  errors are shown by the red dashed lines. The vertical black dotted
  lines mark the transition between $z<3$ and $z>3$.}
\label{fig_evol}
\end{figure}

The one dimensional marginalized constraints are shown in
Figure~\ref{fig12}: the likelihoods obtained in the present work are
represented by the black curves, while the results derived from $z<3$
K07 data set analysed by Viel et al. (2009) are shown as dashed blue
lines. The joint constraints for both PDF measurements are plotted as
the red lines. From this figure we can see that overall there is
agreement between the inferred parameter values for the high redshift
(this work) and low redshift (K07) measurements.  The $1-\sigma$
  parameter constraints from the sample of QSOs analysed in this work
  and for the joint sample including the K07 data are summarized in
  Tab. 5.  The minimum $\chi^2$ for the best fitting model to the PDF
  measured in this work is 52 for 38 degrees of freedom (d.o.f.) which
  is likely to happen 7 per cent of the time.  In contrast, for the
  joint sample the minimum $\chi^2$ is 94 for 83 d.o.f. with has a 21
  per cent probability.  The probability is calculated using the full
  covariance matrix for all the parameters recovered. Note that the
  "degrees of freedom" value does not have a precise meaning if the
  error bars are heavily correlated. However, this happens not to be
  the case in the flux ranges considered here, $F=[0.1$--$0.8]$, where
  correlations are weak.  We have also checked that there is good
  agreement between the derived constraints from the present data set
  when compared to constraints obtained by splitting the sample in two
  and three redshift bins.


\subsection{Implications for the IGM thermal history}

  Finally, in Figure \ref{fig_evol} we display the redshift
  evolution of the temperature at the mean density, $T_{0}$, and the
  slope of the temperature-density relation, $\gamma$, inferred from
  our analysis. We use the broken power-law parameterization around
  $z=3$ introduced previously: $f=f^{\rm A}\times[(1+z)/4]^{f^{\rm
      S}(z)}$, where $f=T_{0}$ or $\gamma$, $f^{\rm A}$ is the
  corresponding normalization factor at $z=3$ and $f^{\rm S}(z)$ is
  the slope of the relation (see Tab. 5). The black solid lines in
  Figure \ref{fig_evol} are the results obtained from this work
  ($z>3$) while the shaded blue regions show their associated
  errors. The red dot-dashed lines are the results obtained from the
  joint sample composed of this work and the K07 data.  The red dashed
  lines show the associated errors. Since the $f^{\rm A}$ and $f^{\rm
    S}(z)$ parameters are correlated and the $f^{\rm S}(z)$ are in
  general poorly constrained, we define the error associated to $f$,
  at a given redshift $z$, as follows:
  \begin{equation}
    \Delta f = \left(f^{\rm A}_{\rm ul}-f^{\rm A}_{\rm ll}\right) \times
    \left(\frac{1+z}{3}\right)^{f^{\rm S}(z)},
  \end{equation}
  where $f^{\rm A}_{\rm ul} = f^{\rm A} + 1\sigma$ and $f^{\rm A}_{\rm
    ll} =f^{\rm A} - 1\sigma$ are, respectively, the $1-\sigma$ upper
  and lower limits on $f^{\rm A}$, and ${f^{\rm S}(z)}$ are the best
  fit values of the slope showed in Tab. 5.

  The temperature at mean density and its redshift evolution, shown
  in the upper panel of Figure \ref{fig_evol}, are not strongly
  constrained by the flux PDF, although our measurements are
  consistent with other recent analyses presented in the literature
  which hint that HeII reionization may be completing around $z\simeq
  3$ (Lidz et al. 2010, Becker et al. 2011).  However, the PDF
  constraints alone are not precise enough to draw conclusions about
  the timing and extent of HeII reionization; the flux PDF is a
  relatively insensitive to $T_{0}$ (see e.g Fig.~\ref{fig_pdf_sim}).
  Alternative measurements from Doppler widths (Schaye et al 2000)
  wavelets (Lidz et al. 2010) or the curvature statistic (Becker et
  al. 2011) are typically better suited for this measurement.

  On the other hand, the flux PDF is sensitive to the slope of the
  temperature-density relation, $\gamma$.  The constraints displayed
  in Fig. \ref{fig_evol}, based on the PDF measurements presented in
  this work, are formally consistent with an isothermal
  temperature-density relation at $z=3$ (i.e. $\gamma \sim 1$),
  although the uncertainties remain large.  A temperature-density
  relation which is close to isothermal may be expected if HeII
  reionization was underway by $z=3$ (e.g. Schaye et al. 2000, Theuns
  et al. 2002).  When combining our data with the K07 measurements at
  lower redshift, however, the inferred $\gamma$ at $z=3$ is shifted
  to lower, inverted $(\gamma<1)$ values, consistent with previous
  results based on the K07 data (e.g. Bolton et al. 2008, Viel et
  al. 2009).  This is because the K07 data is more constraining at
  $z\leq 3$ due to the larger number of QSOs in this sample, and for
  the simple power-law parameterisation we have assumed this also
  pushes our joint constraints to smaller $\gamma$ values at $z\geq
  3$. 

  It has been suggested recently that an inverted
  temperature-density relation at may be explained by an IGM which is
  volumetrically heated by TeV emission from blazars (e.g. Chang et
  al. 2011, Puchwein et al 2011).  This is in contrast to conventional
  photo-ionization heating models which predict a temperature-density
  relation which evolves from isothermal to $\gamma \sim 1.6$
  following reionization (Hui \& Gnedin 1997).  We caution, however,
  the uncertainties on the flux PDF measurements are still large; the
  temperature-density relation suggested by the radiative transfer
  simulations of McQuinn et al. (2011), $\gamma\sim 1.3$ at $z=3$, is 
  compatible at the $1.5\sigma$ level with the value derived from the
  sample considered in this work alone.  The K07 data alone are
  furthermore consistent with an isothermal IGM, again to within
  $1.5\sigma$ (Viel et al. 2009).  We conclude that tighter
  constraints on $\gamma$ from larger data sets, as well as a
  comparison of different measurement techniques to aid with the
  identification possible systematics, will be required for
  distinguishing between different IGM heating scenarios.

\section{Conclusions}

In this paper, we have measured the \lya forest flux PDF from a set of
eight QSO spectra to investigate the thermal state of the IGM at
$z\simeq 3$.  The absorption lines in the Lyman $\alpha$ forest have
been fitted with Voigt profiles, and absorption lines arising from
heavy elements have been carefully identified.  The spectra have then
subsequently had the metal absorption lines removed from within the
\lya forest.  The resulting masked spectra have been used to study the
redshift evolution of the Lyman $\alpha$ forest flux distribution and
the evolution of the HI effective optical depth. Our results can be
summarized as follows.

\begin{itemize}
\item By removing metals from the \lya forest, we attempt to eliminate
  pixels which will introduce a systematic uncertainty into our
  measurement of the flux PDF.  The effect of removing these metals on
  the PDF is to decrease the number of pixels in bins of low flux and
  increase the number of pixels in bins of high flux.  As a result, the
  PDF measured after correcting for metal absorption is slightly
  steeper than the PDF measured without removing metals.  In general,
  the effect of metal removal on the PDF is strongest for flux bins at
  $F<0.6$, in agreement with previous results for other QSO samples at
  lower redshift (K07).\\

\item We measure the PDF for our full sample in two redshift bins,
  both before and after metal decontamination, and find the effect of
  metal absorption on the \lya forest flux PDF is redshift dependent.
  It is more significant at lower redshift, where the average opacity
  of the Lyman $\alpha$ forest is lower and there are fewer absorption
  lines.  This redshift dependence may be understood in terms of the
  larger number of metal lines which are blended with Lyman $\alpha$
  absorption lines toward higher redshift, resulting in the
  identification of fewer isolated metal absorption lines.  On the
  other hand, it is also possible that the metallicity of the
  intergalactic medium decreases with increasing lookback time.\\

\item We present measurements of the effective optical depth,
  $\tau_{\rm eff}$, from our QSO sample, which provides a probe of the
  IGM which is complementary to the PDF.  The values of $\tau_{\rm eff}$
  measured from the QSOs in our sample
  can be fitted with a power-law $\tau_{\rm eff}=0.0028 \pm 0.0003\, 
  (1+z)^{3.45 \pm 0.08}$,
  in agreement with previous measurements by K07 after correcting
  for metal absorption.  
  We also find no substantial evidence for a deviation from a power
  law in the $\log(\tau_{\rm eff})-\log(1+z)$ plane at $z\simeq 3.2$. \\
  There is a weak redshift dependence of the ratio of the HI
  effective optical depth measured before and after removing
  identified metal absorption lines.  The ratio increases toward lower
  redshift, and is consistent with our previous finding from the PDF
  that metal absorption plays a more significant role in the \lya
  forest toward lower redshift.\\

\item We perform a set of high resolution hydrodynamical simulations
  with a range of cosmological and astrophysical parameters and use
  them to extract mock QSO spectra with properties that closely
  resemble those of the observed spectra.  We compute the \lya flux
  PDF and compare it to the observed PDF in the range $F=[0.1$--$0.8]$
  via a Taylor expansion at second order to find deviations from a
  best-guess case (e.g. Viel \& Haehnelt 2006). By implementing this
  method within a Monte Carlo Markov Chain likelihood estimator, we
  present constraints for the IGM thermal state.  At $z=3$, our PDF
  measurements are consistent with an isothermal temperature-density
  relation, $\gamma=0.90\pm0.21$, and a temperature at the mean
  density of $T_{0}=19250 \pm 4800$ K (1$\sigma$ uncertainties).\\

\item We also obtain joint constraints on the IGM thermal state by
  analysing the lower redshift PDF measurements of K07, which is a
  significantly larger data set than the one considered here.  At
  $z=3$, we find the joint data set is consistent with an inverted
  temperature-density relation, $\gamma=0.70\pm 0.12$, with
  $T_{0}=17900\pm 3500$ K (1$\sigma$ uncertainties).\\

\item In conclusion, the present work highlights the role of the flux PDF as
a competitive and quantitative thermometer which may be used to
constraint the IGM temperature-density relation over a wide redshift
range.  The constraints derived from the present work are
complementary with respect to other approaches, such as the wavelets
method (Lidz et al. 2010), the cutoff in the Doppler width-column
density plane (Schaye et al. 2000) or the curvature (Becker et
al. 2011), which are primarily sensitive to higher density regions of
the IGM at $z\simeq 3$.  In the near future, the most promising route
to making further progress on our knowledge of the thermal state of
the IGM will be to carefully compare and contrast results from these
different probes of the IGM temperature-density relation.
\end{itemize}

\section*{Acknowledgments}
FC wishes to thank Eros Vanzella for several 
interesting discussions, Cristian Vignali for useful suggestions 
and Jochen Liske for support in the use of VPGUESS.  
ET acknowledges a fellowship from the
European Commission's Framework Programme 7, through the Marie Curie
Initial Training Network CosmoComp PITN-GA-2009-238356. MV
acknowledges support from INFN/PD-51, ASI-AAE grant a PRIN INAF 2009,
a PRIN MIUR and the European ERC-Starting Grant ``cosmoIGM''.  JSB
acknowledges the support of an ARC Australian postdoctoral fellowship
(DP0984947).  The hydrodynamical simulations used in this work were
performed using the Darwin Supercomputer of the University of
Cambridge High Performance Computing Service
(http://www.hpc.cam.ac.uk/), provided by Dell Inc. using Strategic
Research Infrastructure Funding from the Higher Education Funding
Council for England. Postprocessing of the simulations was performed
at CINECA (Italy) and at the COSMOS supercomputer in Cambridge (UK).

\label{lastpage}

\end{document}